\documentclass[floatfix,twocolumn,prd,nofootinbib,superscriptaddress]{revtex4-2}

\usepackage{color}

\usepackage{dcolumn}
\usepackage{bm}
\usepackage{xspace}
\usepackage{url}

\usepackage{amsmath}
\usepackage{mathrsfs}
\usepackage{graphicx}

\usepackage{bm}
\usepackage{tabularx}
\usepackage{listings}
\usepackage{multirow}
\usepackage{amssymb}
\usepackage{color}
\usepackage{tikz}
\usepackage[mathlines]{lineno}
\usepackage{longtable}
\usepackage{caption}
\usepackage{subcaption}
\DeclareCaptionJustification{fulljustified}{\flushing}
\makeatletter
\expandafter\let\csname longtable*\endcsname\relax
\expandafter\let\csname endlongtable*\endcsname\relax
\makeatother
\usepackage{array}

\newlength\mylength
\newcolumntype{C}[1]{>{\centering\arraybackslash}p{#1}}

\newcommand\apjl{ApJL}

\newcommand\skipme[1]{}

\newcommand{\lt}{\ensuremath <}

\newcommand\optional[1]{}

\def\RIT{Center for Computational Relativity and Gravitation, Rochester Institute of Technology, Rochester, New York
  14623, USA}

\begin{document}

\title{Analysis of GWTC-3 with multiple quasicircular models}
\author{N. Manning}
\affiliation{\RIT}
\author{R. O'Shaughnessy}
\affiliation{\RIT}
\begin{abstract}
The interpretation of gravitational wave sources depends on the specific choice of model used to interpret signals.
Previous analyses of GWTC-3 event candidates using SEOBNRv4PHM, IMRPhenomXPHM, and NRSur7dq4 have sometimes
arrived at notably different conclusions about event properties. Subtle analysis settings and code differences can also
produce notable differences. We revisit 60 publicly available events from the first three observing runs, released
across GWTC-1, GWTC-2, GWTC-2.1, and GWTC-3, using a consistent analysis framework and three waveform
models: two state-of-the-art time-domain models with higher-order modes (SEOBNRv5PHM and IMRPhenomTPHM) and one older
frequency-domain model without higher-order modes (IMRPhenomPv2). The two state-of-the-art models agree well overall,
although about 20\% of events have a visible difference in at least one
one-dimensional marginal posterior. By contrast, the older model often arrives at qualitatively different conclusions
for key events across the mass spectrum. We also identify a bug in published LVK GWTC-2.1 SEOBNRv4PHM results and in
a later reprocessing of GW200105. At the adopted sampling rates, the model's Nyquist-frequency restriction excluded
part of the allowed parameter space and artificially truncated the mass-ratio posterior away from equal mass for seven
low-mass events. These results emphasize the need to use multiple state-of-the-art waveform models and carefully
validated analysis settings to characterize uncertainty in gravitational-wave source properties.
\end{abstract}
\maketitle

\section{Introduction}

The ground-based detector network of the LIGO--Virgo--KAGRA (LVK) Collaboration includes Advanced LIGO
\cite{2015CQGra..32g4001L}, Advanced Virgo
\cite{gw-detectors-Virgo-original-preferred,2015CQGra..32b4001A}, and KAGRA
\cite{2021PTEP.2021eA101A}. These detectors continue to identify coalescing compact binaries
\cite{DiscoveryPaper,LIGO-O1-BBH,LIGO-GW170817-bns,LIGO-O3-NSBH,LIGO-O3-O3b-catalog,LIGO-O3-O3a_final-catalog,LIGO-O4a-cbc-catalog_results,LIGO-O4b-GWTC5}. The cumulative catalog now contains 390 candidate events with
$p_{\rm astro}\geq 0.5$ from the first four observing runs (referred to as O1, O2, O3, and O4)
\cite{LIGO-O4b-GWTC5}. For each candidate event, the GW strain data can be
compared to approximate models for the gravitational radiation emitted from the coalescing (quasicircular) compact
binary, to deduce the distribution of plausible source binary parameters consistent with the data; see, e.g.,   \cite{LIGO-O3-O3b-catalog,LIGO-O4a-cbc-catalog_methods,gwastro-RIFT-Update}  and
references therein.

Without an efficient solution to the two-body problem within general relativity, estimates of the parameters of a
compact binary coalescence (CBC) depend on a culmination of approximate waveform models for the GW produced during the
merger. Previously, GWTC-3 has been analyzed using SEOBNRv4PHM \cite{2018PhRvD..98h4028C,2020PhRvD.102d4055O}
and IMRPhenomXPHM  \cite{gwastro-mergers-IMRPhenomXP}.  In previous analyses these waveform models can
and do conclude differing approximations about individual events’ properties. Both the GWTC-3 catalog and its reanalysis
found several events with notably different properties  \cite{LIGO-O3-O3a-catalog,gwastro-mergers-TousifGWTC3}. Generally, many investigations repeatedly discover
differences between even the best currently-available waveform models, such that the properties of some public events
are notably different when compared with different models \cite{gwastro-Systematics-Williamson2017,gwastro-systematics-ScottFeroz2019,LIGO-O3-O3bcatalog,LIGO-O3-GW190412,LIGO-O3-GW190521-implications,2023arXiv230318046R,gwastro-mergers-TousifGWTC3,LIGO-O4a-cbc-catalog_results}.
Targeted short-author studies of GW200129\_065458 further illustrate that subtle conclusions can depend jointly on waveform
modeling and data treatment \cite{2022Natur.610..652H,2022PhRvD.106j4017P}.

Given the substantial and ongoing need to reassess candidate events with the best available interpretations, frameworks
that enable efficient large-scale reanalysis of events have become vital tools for gravitational wave
astronomy. Previously, an efficient framework for reanalyzing events using asimov
\cite{asimov-paper,gwastro-mergers-rift_asimov_O3-Fernando2024} and RIFT
\cite{gwastro-PENR-RIFT,gwastro-PENR-RIFT-GPU,gwastro-RIFT-Update,gwastro-RIFT_FinerNet} was used to analyze a
subsection of GWTC-3 \cite{LIGO-O3-O3b-catalog}.
In this work, we expand on this earlier analysis by systematically reassessing a selected set of 60 events released
across GWTC-1 through GWTC-3 using three waveform models spanning multiple modeling techniques and eras.

Our analysis presents a comprehensive reanalysis of GWTC-3 using multiple waveform models within a single consistent
framework. The two state-of-the-art models agree well for most events, with a small number of important outliers, while
their broad disagreement with the older model highlights the improvements provided by models that include physics such
as higher-order modes.

This paper is organized as follows. In Section \ref{sec: Methods} we provide an overview of our parameter estimation
approach, the asimov infrastructure, as well as the selection of GWTC-3 events and waveform models. In Section
\ref{sec: results} we provide an overview of our reanalyses of these events. We first describe selected anecdotal
examples to highlight waveform systematics. Then, following previous studies, we present large-scale summary figures and
tables characterizing all the events used. Finally, we compare our reanalyses with published LVK results and discuss the
effect of analysis settings on several exceptional outliers.

\section{Methods}
\label{sec: Methods}
\subsection{Parameter Inference with RIFT}
\label{sec: RIFT}

A coalescing compact binary can be characterized by intrinsic parameters $\bm\lambda$, describing the component
masses and spins, and extrinsic parameters $\bm\theta$, describing the source's spacetime location and orientation
relative to the detectors.

RIFT \cite{gwastro-RIFT_FinerNet,gwastro-PENR-RIFT,gwastro-RIFT-Update} can be understood as a two-stage iterative
process that compares gravitational-wave observations $d$ with predicted signals $h(\bm\lambda,\bm\theta)$. In the first
stage, referred to as ILE, RIFT computes the marginal likelihood
\begin{equation}
 {\cal L}({\bm \lambda})\equiv\int {\cal L}_{\rm full}(\bm{\lambda},\bm\theta)p(\bm\theta)d\bm\theta \, ,
\end{equation}
from the likelihood ${\cal L}_{\rm full}$ of the signal in the multi-detector network, accounting for the detector
response; see \cite{gwastro-PE-AlternativeArchitectures,gwastro-PENR-RIFT} for a more detailed specification. In the
second stage, RIFT interpolates the likelihood information accumulated from previous iterations to approximate
${\cal L}(\bm\lambda)$ and deduce the detector-frame posterior
\begin{equation}
\label{eq:post}
p_{\rm post}(\bm\lambda)=\frac{{\cal L}(\bm{\lambda})p(\bm{\lambda})}
{\int d\bm{\lambda}\,{\cal L}(\bm{\lambda})p(\bm{\lambda})} \, ,
\end{equation}
where $p(\bm\lambda)$ is the prior on intrinsic parameters.

At the end of the iterative calculation, RIFT performs a final pass of ILE on posterior intrinsic draws $\lambda_k$, fairly drawing some (fixed) number of extrinsic parameters $\theta_{k,p}$ from the underlying ILE Monte Carlo integral weighted samples. These combined samples ($\lambda_k,\theta_{k,p}$) provide a full posterior for all intrinsic and extrinsic parameters. Post-processing of each sample provides derived parameters, including the masses of the source frame $m_{i}$.

\subsection{Data, Events, and Settings}
\label{sec: data}

The GWTC-3 catalog reported a cumulative total of 90 compact binary coalescences (CBCs) in the first three
observing runs, incorporating events previously released in GWTC-1, GWTC-2, and GWTC-2.1
\cite{LIGO-O2-Catalog,LIGO-O3-O3a-catalog,LIGO-O3-O3a_final-catalog,LIGO-O3-O3b-catalog}. Of these, 76 events have a false alarm rate $\text{FAR} \lt 1$
$\text{yr}^{-1}$ \cite{LIGO-O3-O3bpop} for at least one pipeline.   Here we further
restrict the analyzed events to only events with $\text{FAR}_{\text{min}} \lt 0.25$  $\text{yr}^{-1}$, consistent with the
events used to assess the global binary black hole population from GWTC-3 \cite{LIGO-O3-O3bpop}, resulting in a
total of 60 analyzed events. Table \ref{tab:events} lists the events we have selected for reanalysis.

All detector strain data and comparison posteriors used in this work are publicly available through the
Gravitational Wave Open Science Center; we use no auxiliary-channel or non-public detector data.

We used the asimov infrastructure \cite{asimov-paper,gwastro-mergers-rift_asimov_O3-Fernando2024} to configure our
analysis settings as consistently as possible with prior work. Asimov provides a code-neutral framework for encoding
event-specific settings and efficiently repeating inferences with multiple waveform models. Together with RIFT, this
infrastructure makes the analysis reproducible at the scale of many events and waveform choices.
Our production configuration also uses RIFT's adaptive-volume integrator \cite{Tiwari_2023}, one of the sampling
improvements incorporated into the current RIFT workflow.
Following prior RIFT studies, this operating point was checked with end-to-end probability--probability tests using
randomly drawn sources \cite{gwastro-PENR-RIFT,gwastro-RIFT-Update}.

Specifically,  our analyses employ nearly identical settings to the production analyses employed for new events first
described GWTC-3 \cite{LIGO-O3-O3b-catalog}.  To be consistent with their final reweighted results, we adopt a distance
prior which is uniform in per unit comoving volume per unit detector frame time (i.e., a volume element $(1+z)^{-1}dV_c$).   For technical reasons -- RIFT internally uses discrete time, so events with narrow event time posteriors could
require a high sampling rate -- we have chosen to adopt a higher sampling rate of $4096$ Hz or larger
for all events, also necessitating regeneration of the estimated noise PSD.  As a result, our analyses adopt extremely
similar but not absolutely identical analysis settings as previously published results.

As regards RIFT internal settings, we adopt pipeline-default settings chosen when RIFT and asimov operate together,
notably including the internal intrinsic threshold $n_{\rm eff}=10$ for each intrinsic Monte Carlo integral.

As an application of our framework, we revisit several of the analyses presented in GWTC-3, emphasizing events
previously analyzed using RIFT with \mbox{SEOBNRv4PHM} \cite{2018PhRvD..98h4028C,2020PhRvD.102d4055O}.  The previously
published RIFT GWTC-3 analyses provided extremely few posterior samples. To demonstrate the utility of our framework
for reproducible investigation of waveform systematics, we perform new analyses with three waveform models:
\mbox{IMRPhenomPv2} \cite{gwastro-mergers-IMRPhenomP}, \mbox{IMRPhenomTPHM}  \cite{2022PhRvD.105h4040E}, \mbox{SEOBNRv5PHM} \cite{2023arXiv230318046R,2023arXiv230318203M}.

These choices span both modeling technique and generation. IMRPhenomPv2 is an older frequency-domain phenomenological,
single-precessing-spin model without higher-order modes, whereas IMRPhenomTPHM is a time-domain phenomenological model
with precession and higher-order modes. SEOBNRv5PHM instead follows the time-domain effective-one-body approach through
inspiral, merger, and ringdown, also including precession and higher-order modes. Thus, the controlled comparison places
two time-domain, higher-order-mode models against one frequency-domain model without higher-order modes, while comparing
the two state-of-the-art models separately isolates current cross-family modeling differences.

    \setlength\tabcolsep{3pt}
    \begin{longtable*}[c]{l c c c c c c c c c c}
    \caption{A table showing the total events analyzed. The columns contain the minimum FAR \cite{LIGO-O3-O3a_final-catalog, LIGO-O3-O3b-catalog, LIGO-O3-O3bpop}, key source parameters, starting frequency, sampling rate, and natural log ($\ln$) of the Bayes factor associated with each event analysis. The values represented in each row correspond to the analysis (SEOBNRv5PHM or IMRPhenomTPHM) that resulted in the highest $\ln$ Bayes factor. The inferred properties provide the median values and 90\% confidence intervals.}\label{tab:events}
    \\
    \hline \hline
    Name & FAR$_{\mathrm{min}}$\ & $\mathcal{M}_c$ & $m_1$ & $m_2$ & $\chi_{\mathrm{eff}}$ & $\chi_{\mathrm{p}}$ & $f_{\mathrm{start}}$\ & $s$\ & $\ln \mathcal{B}$ & Approx.\\
    & [yr$^{-1}$] & [$M_{\odot}$] & [$M_{\odot}$] & [$M_{\odot}$] & & & [Hz] & [Hz] & & \\
    \hline
    \endfirsthead
    \caption{\it{(Continued)}}
    \\
    \hline \hline
    Name & FAR$_{\mathrm{min}}$\ & $\mathcal{M}_c$ & $m_1$ & $m_2$ & $\chi_{\mathrm{eff}}$ & $\chi_{\mathrm{p}}$ & $f_{\mathrm{start}}$\ & $s$\ & $\ln \mathcal{B}$ & Approx.\\
    & [yr$^{-1}$] & [$M_{\odot}$] & [$M_{\odot}$] & [$M_{\odot}$] & & & [Hz] & [Hz] & & \\
    \hline
    \endhead
    \hline
    \endfoot
    \hline
    \endlastfoot

 GW170104 & $<1\times10^{-5}$ & $25.57_{-1.07}^{+0.99}$ & $34.91_{-3.63}^{+4.67}$ & $25.00_{-3.46}^{+3.17}$ & $-0.05_{-0.12}^{+0.10}$ & $0.38_{-0.21}^{+0.28}$ & 20 & 4096 & 2.12 & SEOB\\

 GW170608 & $<1\times10^{-5}$ & $8.49_{-0.03}^{+0.03}$ & $11.11_{-0.94}^{+1.55}$ & $8.59_{-0.97}^{+0.77}$ & $0.04_{-0.03}^{+0.04}$ & $0.27_{-0.15}^{+0.23}$ & 20 & 4096 & 1.79 & SEOB\\

 GW170729 & $1.8\times10^{-1}$ & $49.40_{-6.60}^{+6.13}$ & $77.13_{-9.89}^{+11.87}$ & $42.57_{-10.41}^{+11.47}$ & $0.27_{-0.20}^{+0.17}$ & $0.39_{-0.20}^{+0.25}$ & 20 & 4096 & 2.54 & IMRP\\

 GW170809 & $<1\times10^{-5}$ & $29.83_{-1.15}^{+1.43}$ & $40.05_{-3.94}^{+5.56}$ & $29.89_{-4.22}^{+3.40}$ & $0.06_{-0.10}^{+0.12}$ & $0.37_{-0.20}^{+0.27}$ & 20 & 4096 & 2.09 & IMRP\\

 GW170814 & $<1\times10^{-5}$ & $27.29_{-0.83}^{+0.87}$ & $35.44_{-2.76}^{+3.98}$ & $27.89_{-2.97}^{+2.51}$ & $0.11_{-0.08}^{+0.08}$ & $0.52_{-0.27}^{+0.25}$ & 20 & 4096 & 2.33 & IMRP\\

 GW170818 & $<1\times10^{-5}$ & $32.39_{-1.54}^{+1.57}$ & $42.48_{-3.62}^{+5.04}$ & $32.90_{-4.15}^{+3.50}$ & $-0.06_{-0.13}^{+0.12}$ & $0.50_{-0.25}^{+0.26}$ & 20 & 4096 & 2.70 & IMRP\\

 GW170823 & $<1\times10^{-5}$ & $38.94_{-2.73}^{+3.08}$ & $52.03_{-5.49}^{+7.28}$ & $39.34_{-6.29}^{+4.99}$ & $0.06_{-0.13}^{+0.14}$ & $0.43_{-0.23}^{+0.27}$ & 20 & 4096 & 2.38 & IMRP\\

 GW190408\_181802 & $<1\times10^{-5}$ & $24.00_{-0.76}^{+0.76}$ & $32.03_{-3.02}^{+4.24}$ & $23.94_{-3.09}^{+2.64}$ & $-0.01_{-0.10}^{+0.09}$ & $0.38_{-0.21}^{+0.28}$ & 20 & 4096 & 1.86 & IMRP\\

 GW190412\_053044 & $<1\times10^{-5}$ & $15.19_{-0.08}^{+0.07}$ & $35.51_{-2.32}^{+2.61}$ & $9.32_{-0.49}^{+0.51}$ & $0.25_{-0.04}^{+0.04}$ & $0.24_{-0.08}^{+0.09}$ & 20 & 4096 & 6.65 & SEOB\\

 GW190413\_134308 & $1.81\times10^{-1}$ & $58.07_{-6.69}^{+6.40}$ & $80.98_{-10.05}^{+12.40}$ & $56.85_{-13.50}^{+10.18}$ & $0.02_{-0.17}^{+0.17}$ & $0.44_{-0.24}^{+0.28}$ & 10 & 4096 & 2.44 & SEOB\\

 GW190421\_213856 & $2.83\times10^{-3}$ & $45.75_{-3.86}^{+3.70}$ & $61.93_{-6.25}^{+7.97}$ & $45.72_{-8.59}^{+6.73}$ & $-0.09_{-0.16}^{+0.14}$ & $0.44_{-0.23}^{+0.27}$ & 16 & 4096 & 2.41 & SEOB\\

 GW190503\_185404 & $<1\times10^{-5}$ & $37.88_{-3.74}^{+3.22}$ & $52.59_{-6.34}^{+7.55}$ & $36.68_{-7.11}^{+6.11}$ & $-0.04_{-0.17}^{+0.13}$ & $0.42_{-0.23}^{+0.29}$ & 20 & 4096 & 2.25 & SEOB\\

 GW190512\_180714 & $<1\times10^{-5}$ & $18.67_{-0.30}^{+0.34}$ & $28.65_{-4.51}^{+4.46}$ & $16.32_{-2.10}^{+3.03}$ & $0.03_{-0.07}^{+0.08}$ & $0.26_{-0.15}^{+0.23}$ & 20 & 4096 & 1.55 & IMRP\\

 GW190513\_205428 & $<1\times10^{-5}$ & $29.97_{-2.17}^{+4.16}$ & $49.03_{-8.89}^{+9.90}$ & $25.99_{-5.66}^{+6.86}$ & $0.14_{-0.15}^{+0.21}$ & $0.33_{-0.18}^{+0.26}$ & 20 & 4096 & 2.17 & IMRP\\

 GW190517\_055101 & $3.47\times10^{-4}$ & $36.75_{-1.99}^{+2.14}$ & $51.54_{-6.03}^{+7.81}$ & $35.14_{-5.40}^{+5.44}$ & $0.57_{-0.10}^{+0.11}$ & $0.46_{-0.17}^{+0.17}$ & 20 & 4096 & 7.24 & IMRP\\

 GW190519\_153544 & $<1\times10^{-5}$ & $66.15_{-5.33}^{+4.60}$ & $95.29_{-7.76}^{+8.62}$ & $61.11_{-9.32}^{+9.04}$ & $0.31_{-0.12}^{+0.12}$ & $0.45_{-0.20}^{+0.24}$ & 10 & 4096 & 4.78 & IMRP\\

 GW190521\_030229 & $<1\times10^{-5}$ & $107.99_{-11.16}^{+8.73}$ & $148.44_{-13.25}^{+15.29}$ & $104.62_{-20.46}^{+17.47}$ & $-0.08_{-0.21}^{+0.17}$ & $0.55_{-0.26}^{+0.24}$ & 10 & 4096 & 2.50 & IMRP\\

 GW190521\_074359 & $1.0\times10^{-2}$ & $39.96_{-1.45}^{+1.56}$ & $52.49_{-3.76}^{+4.33}$ & $40.54_{-4.08}^{+3.89}$ & $0.11_{-0.08}^{+0.08}$ & $0.38_{-0.20}^{+0.24}$ & 20 & 4096 & 2.48 & IMRP\\

 GW190527\_092055 & $2.28\times10^{-1}$ & $35.01_{-4.17}^{+6.12}$ & $55.36_{-10.16}^{+17.93}$ & $32.05_{-9.89}^{+9.42}$ & $0.12_{-0.18}^{+0.19}$ & $0.43_{-0.23}^{+0.29}$ & 10 & 4096 & 2.26 & IMRP\\

 GW190602\_175927 & $<1\times10^{-5}$ & $74.37_{-7.91}^{+7.23}$ & $103.77_{-11.20}^{+12.77}$ & $72.83_{-16.77}^{+12.98}$ & $0.13_{-0.15}^{+0.16}$ & $0.42_{-0.22}^{+0.28}$ & 10 & 4096 & 2.52 & SEOB\\

 GW190620\_030421 & $1.12\times10^{-2}$ & $57.79_{-6.32}^{+5.99}$ & $86.17_{-11.71}^{+13.01}$ & $52.58_{-11.56}^{+11.34}$ & $0.37_{-0.15}^{+0.13}$ & $0.48_{-0.21}^{+0.22}$ & 10 & 4096 & 4.61 & SEOB\\

 GW190630\_185205 & $<1\times10^{-5}$ & $28.96_{-0.95}^{+1.07}$ & $41.81_{-5.30}^{+6.09}$ & $26.99_{-3.96}^{+4.36}$ & $0.07_{-0.08}^{+0.09}$ & $0.29_{-0.15}^{+0.22}$ & 20 & 4096 & 1.90 & IMRP\\

 GW190701\_203306 & $5.71\times10^{-3}$ & $54.69_{-4.77}^{+4.56}$ & $73.01_{-7.14}^{+8.85}$ & $55.38_{-10.15}^{+7.78}$ & $-0.09_{-0.18}^{+0.15}$ & $0.45_{-0.23}^{+0.27}$ & 10 & 4096 & 2.63 & SEOB\\

 GW190706\_222641 & $<1\times10^{-5}$ & $78.71_{-8.86}^{+6.84}$ & $110.62_{-11.66}^{+13.91}$ & $75.53_{-18.38}^{+13.88}$ & $0.31_{-0.19}^{+0.16}$ & $0.44_{-0.20}^{+0.24}$ & 10 & 4096 & 3.44 & IMRP\\

 GW190707\_093326 & $<1\times10^{-5}$ & $9.89_{-0.06}^{+0.08}$ & $13.40_{-1.40}^{+1.83}$ & $9.70_{-1.09}^{+1.07}$ & $-0.06_{-0.05}^{+0.06}$ & $0.26_{-0.15}^{+0.22}$ & 20 & 4096 & 1.87 & SEOB\\

 GW190708\_232457 & $3.09\times10^{-4}$ & $15.48_{-0.13}^{+0.13}$ & $20.42_{-1.86}^{+3.01}$ & $15.56_{-1.92}^{+1.53}$ & $0.02_{-0.05}^{+0.05}$ & $0.32_{-0.19}^{+0.28}$ & 20 & 4096 & 1.44 & IMRP\\

 GW190720\_000836 & $<1\times10^{-5}$ & $10.35_{-0.06}^{+0.06}$ & $14.64_{-1.97}^{+3.42}$ & $9.73_{-1.63}^{+1.40}$ & $0.14_{-0.06}^{+0.08}$ & $0.35_{-0.17}^{+0.24}$ & 20 & 4096 & 4.29 & IMRP\\

 GW190727\_060333 & $<1\times10^{-5}$ & $44.78_{-3.95}^{+3.72}$ & $59.43_{-5.90}^{+8.03}$ & $45.59_{-8.41}^{+5.94}$ & $0.11_{-0.18}^{+0.21}$ & $0.46_{-0.23}^{+0.26}$ & 10 & 4096 & 2.83 & IMRP\\

 GW190728\_064510 & $<1\times10^{-5}$ & $10.13_{-0.05}^{+0.05}$ & $14.10_{-1.79}^{+3.18}$ & $9.66_{-1.56}^{+1.33}$ & $0.12_{-0.05}^{+0.09}$ & $0.33_{-0.16}^{+0.24}$ & 20 & 4096 & 3.85 & IMRP\\

 GW190803\_022701 & $7.32\times10^{-2}$ & $43.01_{-3.36}^{+3.43}$ & $57.85_{-6.00}^{+7.88}$ & $43.09_{-7.56}^{+5.95}$ & $-0.02_{-0.16}^{+0.14}$ & $0.43_{-0.23}^{+0.28}$ & 20 & 4096 & 2.58 & SEOB\\

 GW190814\_211039 & $<1\times10^{-5}$ & $6.41_{-0.01}^{+0.01}$ & $24.04_{-0.88}^{+0.60}$ & $2.75_{-0.05}^{+0.07}$ & $-0.02_{-0.05}^{+0.03}$ & $0.04_{-0.02}^{+0.03}$ & 20 & 8192 & 0.23 & SEOB\\

 GW190828\_063405 & $<1\times10^{-5}$ & $35.05_{-1.92}^{+1.95}$ & $45.31_{-3.87}^{+5.09}$ & $36.09_{-4.15}^{+3.51}$ & $0.22_{-0.11}^{+0.10}$ & $0.48_{-0.23}^{+0.24}$ & 20 & 4096 & 3.33 & IMRP\\

 GW190828\_065509 & $<1\times10^{-5}$ & $17.42_{-0.42}^{+0.39}$ & $30.35_{-5.33}^{+5.20}$ & $13.59_{-1.86}^{+2.76}$ & $0.06_{-0.10}^{+0.10}$ & $0.28_{-0.15}^{+0.24}$ & 20 & 4096 & 2.12 & SEOB\\

 GW190910\_112807 & $2.87\times10^{-3}$ & $43.17_{-2.40}^{+2.44}$ & $56.59_{-4.56}^{+5.27}$ & $44.05_{-5.62}^{+4.93}$ & $-0.03_{-0.12}^{+0.11}$ & $0.41_{-0.23}^{+0.27}$ & 20 & 4096 & 2.07 & SEOB\\

 GW190915\_235702 & $<1\times10^{-5}$ & $32.17_{-1.76}^{+1.72}$ & $43.48_{-4.53}^{+6.25}$ & $31.78_{-4.70}^{+3.97}$ & $-0.03_{-0.14}^{+0.12}$ & $0.51_{-0.26}^{+0.27}$ & 20 & 4096 & 2.78 & IMRP\\

 GW190924\_021846 & $<1\times10^{-5}$ & $6.44_{-0.02}^{+0.02}$ & $9.70_{-1.60}^{+2.47}$ & $5.72_{-1.01}^{+1.04}$ & $0.03_{-0.06}^{+0.12}$ & $0.24_{-0.13}^{+0.22}$ & 20 & 8192 & 1.78 & IMRP\\

 GW190925\_232845 & $7.2\times10^{-3}$ & $18.61_{-0.47}^{+0.51}$ & $25.18_{-2.60}^{+4.50}$ & $18.31_{-2.74}^{+2.14}$ & $0.10_{-0.10}^{+0.10}$ & $0.41_{-0.22}^{+0.28}$ & 20 & 4096 & 2.36 & IMRP\\

 GW190929\_012149 & $1.55\times10^{-1}$ & $54.21_{-8.58}^{+9.71}$ & $101.37_{-12.61}^{+13.70}$ & $39.83_{-11.51}^{+17.05}$ & $0.00_{-0.14}^{+0.13}$ & $0.30_{-0.17}^{+0.30}$ & 10 & 4096 & 1.96 & IMRP\\

 GW190930\_133541 & $1.23\times10^{-2}$ & $9.84_{-0.13}^{+0.09}$ & $13.46_{-1.56}^{+2.82}$ & $9.51_{-1.49}^{+1.19}$ & $0.13_{-0.08}^{+0.08}$ & $0.35_{-0.18}^{+0.25}$ & 20 & 4096 & 3.00 & SEOB\\

 GW191105\_143521 & $1.18\times10^{-2}$ & $9.57_{-0.06}^{+0.07}$ & $12.89_{-1.33}^{+2.25}$ & $9.43_{-1.30}^{+1.06}$ & $-0.03_{-0.05}^{+0.06}$ & $0.29_{-0.17}^{+0.25}$ & 20 & 8192 & 1.51 & SEOB\\

 GW191109\_010717 & $1.8\times10^{-4}$ & $59.14_{-4.03}^{+3.51}$ & $80.20_{-6.00}^{+6.31}$ & $58.02_{-7.99}^{+7.67}$ & $-0.37_{-0.12}^{+0.12}$ & $0.57_{-0.21}^{+0.19}$ & 20 & 4096 & 6.69 & IMRP\\

 GW191127\_050227 & $2.49\times10^{-1}$ & $50.44_{-11.07}^{+10.70}$ & $79.87_{-17.50}^{+21.18}$ & $43.64_{-14.15}^{+15.50}$ & $0.16_{-0.20}^{+0.21}$ & $0.46_{-0.25}^{+0.28}$ & 20 & 4096 & 2.57 & IMRP\\

 GW191129\_134029 & $<1\times10^{-5}$ & $8.48_{-0.03}^{+0.04}$ & $11.44_{-1.23}^{+2.47}$ & $8.33_{-1.32}^{+0.96}$ & $0.04_{-0.04}^{+0.08}$ & $0.30_{-0.17}^{+0.25}$ & 20 & 4096 & 1.78 & IMRP\\

 GW191204\_171526 & $<1\times10^{-5}$ & $9.70_{-0.03}^{+0.03}$ & $12.80_{-1.13}^{+1.57}$ & $9.74_{-0.99}^{+0.91}$ & $0.15_{-0.02}^{+0.03}$ & $0.33_{-0.14}^{+0.18}$ & 20 & 4096 & 4.83 & SEOB\\

 GW191216\_213338 & $<1\times10^{-5}$ & $8.93_{-0.03}^{+0.03}$ & $12.65_{-1.61}^{+2.01}$ & $8.39_{-1.02}^{+1.14}$ & $0.10_{-0.04}^{+0.06}$ & $0.25_{-0.12}^{+0.17}$ & 20 & 4096 & 3.50 & SEOB\\

 GW191222\_033537 & $<1\times10^{-5}$ & $50.47_{-4.09}^{+4.38}$ & $67.49_{-6.97}^{+8.81}$ & $51.23_{-8.89}^{+7.15}$ & $-0.07_{-0.15}^{+0.13}$ & $0.44_{-0.23}^{+0.28}$ & 20 & 4096 & 2.18 & SEOB\\

 GW191230\_180458 & $5.02\times10^{-2}$ & $62.41_{-5.41}^{+5.25}$ & $82.52_{-8.16}^{+9.98}$ & $63.63_{-11.18}^{+8.43}$ & $-0.06_{-0.17}^{+0.15}$ & $0.46_{-0.24}^{+0.27}$ & 10 & 4096 & 2.71 & SEOB\\

 GW200105\_162426 & $2.04\times10^{-1}$ & $3.62_{-0.00}^{+0.00}$ & $9.31_{-1.00}^{+0.60}$ & $2.06_{-0.09}^{+0.19}$ & $-0.03_{-0.10}^{+0.06}$ & $0.11_{-0.06}^{+0.07}$ & 20 & 8192 & 0.84 & IMRP\\

 GW200112\_155838 & $<1\times10^{-5}$ & $33.54_{-1.50}^{+1.80}$ & $44.72_{-4.18}^{+5.22}$ & $33.89_{-4.49}^{+3.79}$ & $0.04_{-0.10}^{+0.11}$ & $0.38_{-0.21}^{+0.27}$ & 20 & 4096 & 2.01 & SEOB\\

 GW200128\_022011 & $4.29\times10^{-3}$ & $50.45_{-3.62}^{+3.94}$ & $66.58_{-6.76}^{+9.22}$ & $51.05_{-7.26}^{+6.25}$ & $0.15_{-0.14}^{+0.13}$ & $0.54_{-0.25}^{+0.23}$ & 20 & 4096 & 2.76 & IMRP\\

 GW200129\_065458 & $<1\times10^{-5}$ & $32.07_{-0.95}^{+0.95}$ & $39.85_{-2.16}^{+3.16}$ & $34.20_{-2.79}^{+2.12}$ & $0.10_{-0.07}^{+0.07}$ & $0.50_{-0.24}^{+0.25}$ & 20 & 4096 & 3.03 & SEOB\\

 GW200202\_154313 & $<1\times10^{-5}$ & $8.14_{-0.03}^{+0.03}$ & $11.32_{-0.90}^{+1.40}$ & $7.78_{-0.78}^{+0.62}$ & $0.04_{-0.04}^{+0.05}$ & $0.24_{-0.13}^{+0.20}$ & 20 & 4096 & 1.46 & SEOB\\

 GW200208\_130117 & $3.11\times10^{-4}$ & $39.15_{-3.10}^{+3.09}$ & $53.35_{-5.72}^{+7.15}$ & $38.74_{-6.88}^{+5.74}$ & $-0.07_{-0.15}^{+0.14}$ & $0.40_{-0.21}^{+0.28}$ & 20 & 4096 & 2.41 & SEOB\\

 GW200209\_085452 & $4.64\times10^{-2}$ & $41.71_{-4.35}^{+4.28}$ & $54.99_{-6.08}^{+7.94}$ & $42.41_{-7.50}^{+6.17}$ & $-0.09_{-0.17}^{+0.14}$ & $0.48_{-0.25}^{+0.27}$ & 20 & 4096 & 2.84 & IMRP\\

 GW200219\_094415 & $9.94\times10^{-4}$ & $42.93_{-3.85}^{+3.56}$ & $58.78_{-6.31}^{+8.13}$ & $42.49_{-8.74}^{+6.53}$ & $-0.09_{-0.17}^{+0.14}$ & $0.44_{-0.24}^{+0.28}$ & 20 & 4096 & 2.53 & SEOB\\

 GW200224\_222234 & $<1\times10^{-5}$ & $41.70_{-7.00}^{+2.95}$ & $54.49_{-3.90}^{+4.83}$ & $42.59_{-13.74}^{+5.67}$ & $0.05_{-0.11}^{+0.12}$ & $0.40_{-0.24}^{+0.32}$ & 20 & 4096 & 1.81 & IMRP\\

 GW200225\_060421 & $<1\times10^{-5}$ & $17.94_{-0.88}^{+0.62}$ & $23.46_{-2.00}^{+2.89}$ & $18.06_{-2.31}^{+1.87}$ & $-0.08_{-0.14}^{+0.11}$ & $0.49_{-0.25}^{+0.26}$ & 20 & 4096 & 2.67 & SEOB\\

 GW200302\_015811 & $1.12\times10^{-1}$ & $29.23_{-1.98}^{+3.24}$ & $48.72_{-6.57}^{+6.71}$ & $24.09_{-4.20}^{+7.17}$ & $0.01_{-0.13}^{+0.14}$ & $0.30_{-0.17}^{+0.27}$ & 20 & 4096 & 1.86 & IMRP\\

 GW200311\_115853 & $<1\times10^{-5}$ & $31.79_{-1.47}^{+1.48}$ & $41.25_{-3.29}^{+4.66}$ & $32.77_{-4.25}^{+3.19}$ & $-0.08_{-0.11}^{+0.10}$ & $0.43_{-0.22}^{+0.27}$ & 20 & 4096 & 2.42 & SEOB\\

 GW200316\_215756 & $<1\times10^{-5}$ & $10.68_{-0.07}^{+0.07}$ & $15.38_{-2.23}^{+4.07}$ & $9.86_{-1.81}^{+1.56}$ & $0.11_{-0.06}^{+0.10}$ & $0.31_{-0.15}^{+0.23}$ & 20 & 4096 & 3.51 & SEOB\\
    \end{longtable*}
    \setlength\tabcolsep{6pt}

\subsection{Diagnostics}

We use the Jensen--Shannon (JS) divergence \cite{JSD_paper} to quantify differences between one-dimensional marginalized
posteriors obtained with our waveform models and between our results and the public LVK posteriors distributed through
the Gravitational Wave Open Science Center
\cite{LIGO-O3-O3a_final-catalog,LIGO-O3-O3b-catalog,2023ApJS..267...29A}. The JS divergence between probability density functions $p(x)$
and $q(x)$ is
  \begin{align}
    JSD(p,q) = \frac{KLD(p\parallel m) + KLD(q \parallel m)}{2} \; ,
  \end{align}
and is a symmetrized extension of the Kullback--Leibler divergence \cite{KLD_paper}, where
$m(x)=[p(x)+q(x)]/2$ and
  \begin{align}
    KLD(p \parallel q) = \int p(x) \log_2{\frac{p(x)}{q(x)}}dx\;,
  \end{align}
is the Kullback--Leibler divergence. A JS divergence of 0 signifies identical distributions, while a value of 1
corresponds to distributions with disjoint support. Following the convention used in the thesis analysis, we use
${\rm JSD}=0.015$ as a practical threshold above which differences in the marginal posteriors are generally visible by eye.

The evidence $Z = \int {\cal L}(\bm{\theta})p(\bm{\theta})d\bm{\theta}$ quantifies how well a model fits the
observation. Comparing evidences provides a complementary global diagnostic that retains parameter correlations; large
differences indicate that the data prefer one waveform model over another.

\section{Results}
\label{sec: results}

\subsection{Summary statistics and outlier identification}
\begin{figure}[h]

\includegraphics[width=\columnwidth]{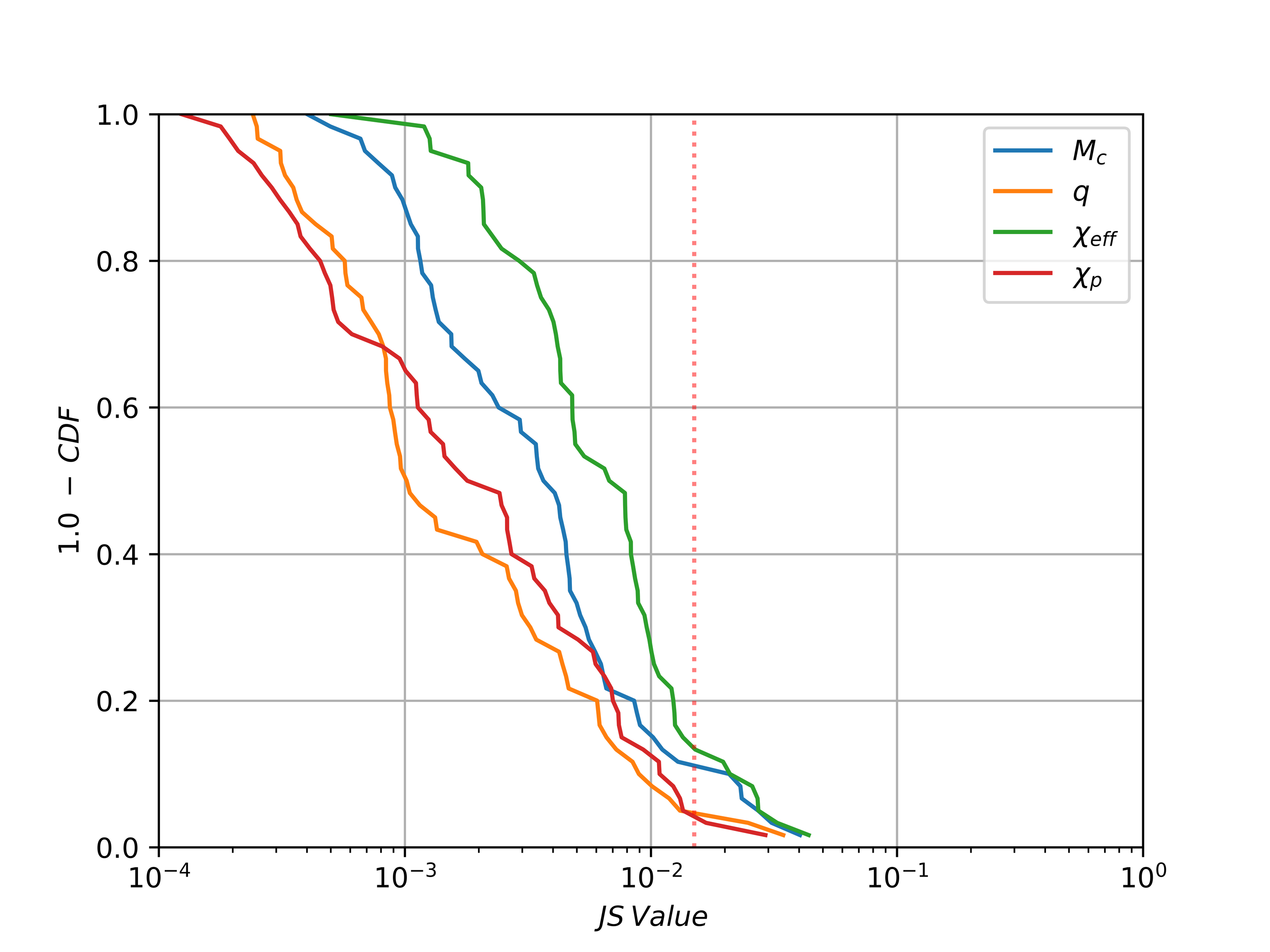}
\caption{ \label{fig:CDF}  Inverse CDF of the Jensen-Shannon Divergence between SEOBNRv5PHM
  and IMRPhenomTPHM posteriors generated in this work,
  for the source frame chirp mass $M_{c}$, mass ratio $q$, effective inspiral spin parameter $\chi_{eff}$, and spin
  precession parameter $\chi_{p}$ over all events analyzed in this work. A significant fraction of events show
  noteworthy disagreement $>10^{-2}$ on one or more one-dimensional marginal distributions.}
\end{figure}

Figure \ref{fig:CDF} shows a simple summary statistic---the Jensen--Shannon (JS) divergence---quantifying differences
between marginal one-dimensional posterior distributions for each event. The JS divergence was calculated between each
waveform over the indicated intrinsic parameters. The figure shows the fraction of events with JS divergence below the
threshold indicated on the horizontal axis.
For most events, this figure demonstrates that parameter inferences derived using the two most sophisticated waveforms
in our analysis suite (IMRPhenomTPHM, SEOBNRv5PHM) arrive at similar though slightly different conclusions about all key
parameters (i.e., the JS divergence is between $10^{-3}$ and $10^{-2}$ -- above the threshold for statistically
indistinguishable but below the threshold where differences in one-dimensional distributions are concerning to the
human eye).

Complete event-by-event and model-by-model comparisons are provided in the first author's thesis
\cite{Manning_2026}. Here we retain only the figures needed to establish the catalog-level trends and motivate the
cross-comparison with published LVK results.

\subsection{Detailed posterior comparisons: Outliers}

\begin{figure*}[h]

  \includegraphics[width=\linewidth]{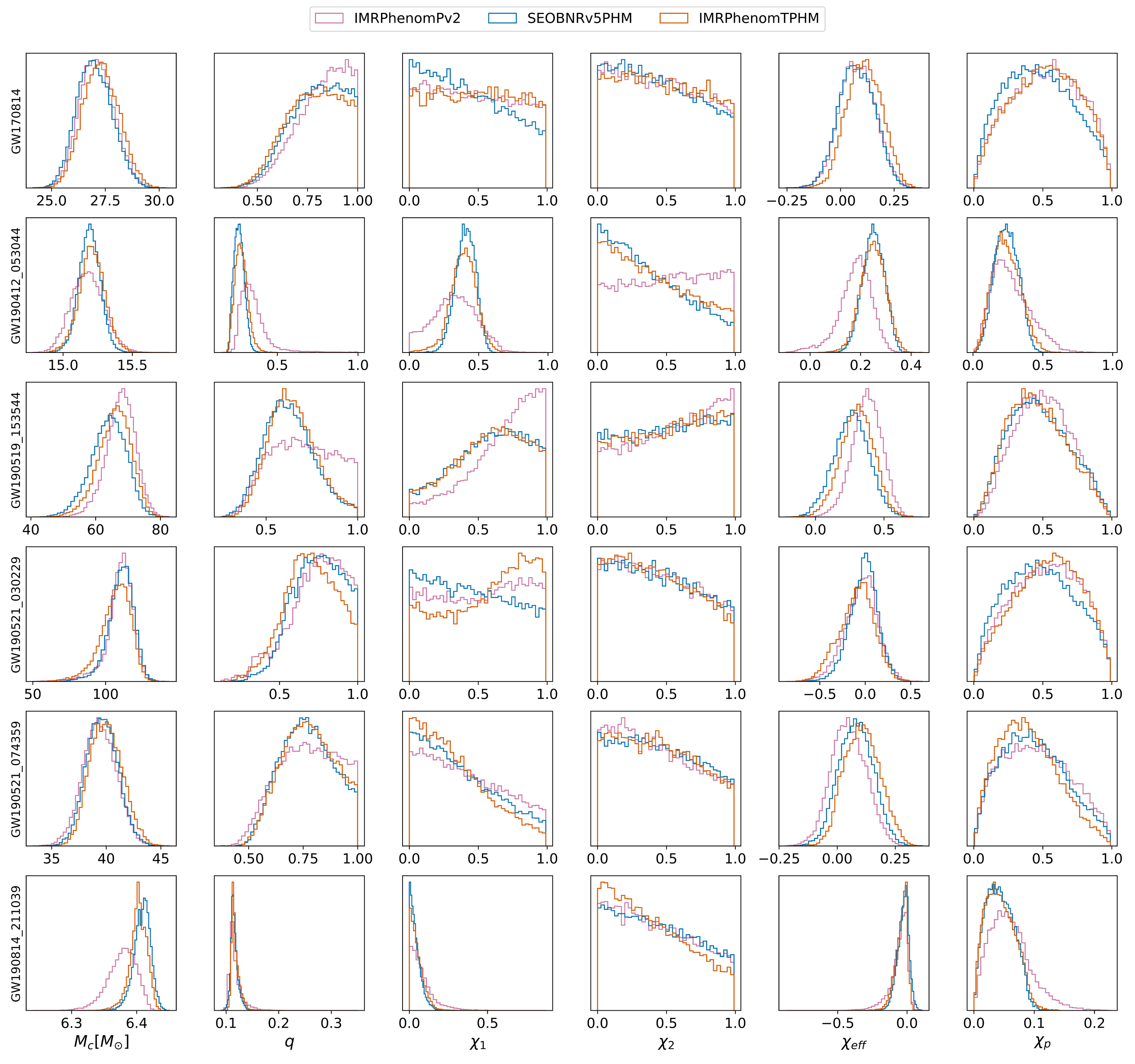}

\caption{ \label{fig:posteriors} Posteriors for the chirp mass $M_{c}$, mass ratio $q$, spin magnitudes $\chi_{1}$ and $\chi_{2}$, effective inspiral spin parameter $\chi_{eff}$, and spin precession parameter $\chi_{p}$ for selected events that have the most significant differences between results obtained using IMRPhenomPv2 (pink), SEOBNRv5PHM (blue), IMRPhenomTPHM (orange).}
\end{figure*}
\addtocounter{figure}{-1}
\begin{figure*}[h]

  \includegraphics[width=\linewidth]{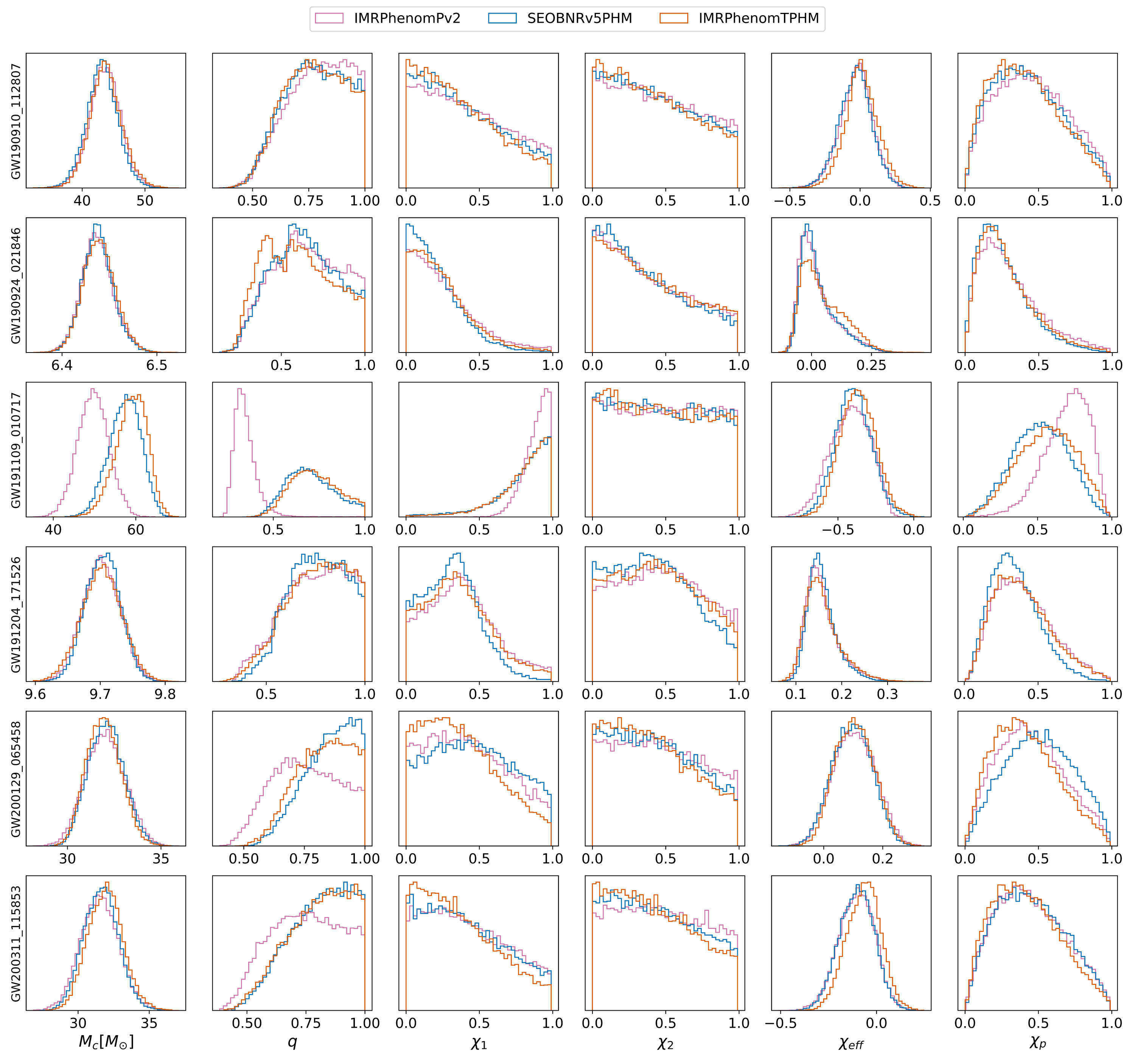}

\caption{ \label{fig:posteriors_part_2} (cont.) Posteriors for the chirp mass $M_{c}$, mass ratio $q$, spin magnitudes $\chi_{1}$ and $\chi_{2}$, effective inspiral spin parameter $\chi_{eff}$, and spin precession parameter $\chi_{p}$ for selected events that have the most significant differences between results obtained using IMRPhenomPv2 (pink), SEOBNRv5PHM (blue), IMRPhenomTPHM (orange).}
\end{figure*}

Figure \ref{fig:posteriors} shows marginal posteriors for selected outliers. IMRPhenomPv2 is occasionally adequate, but
often differs substantially from both state-of-the-art models. GW190412\_053044 exemplifies the importance of higher-order modes
for extracting the information in the signal \cite{LIGO-O3-GW190412}: only the newer models provide strong evidence for
nonzero primary spin. For GW190521\_030229 and GW190814\_211039, the older model also produces qualitatively different
conclusions about the mass ratio and chirp mass, respectively; the dedicated LIGO--Virgo analysis of GW190814 likewise
demonstrated the importance of higher-order modes for this highly asymmetric source \cite{2020ApJ...896L..44A}.

\begin{figure*}
    \centering
    \includegraphics[width=0.32\linewidth]{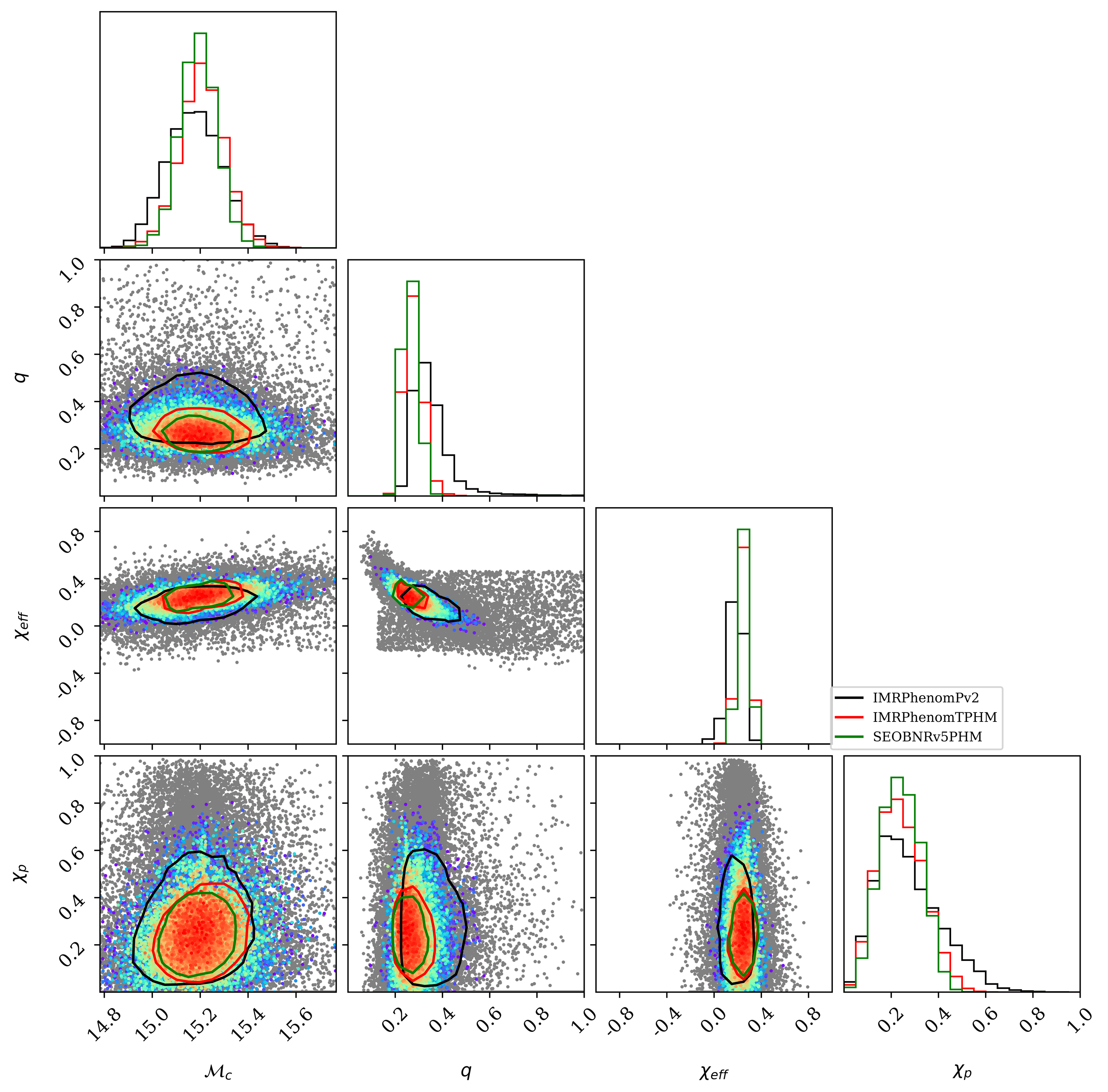}
    \hfill
    \includegraphics[width=0.32\linewidth]{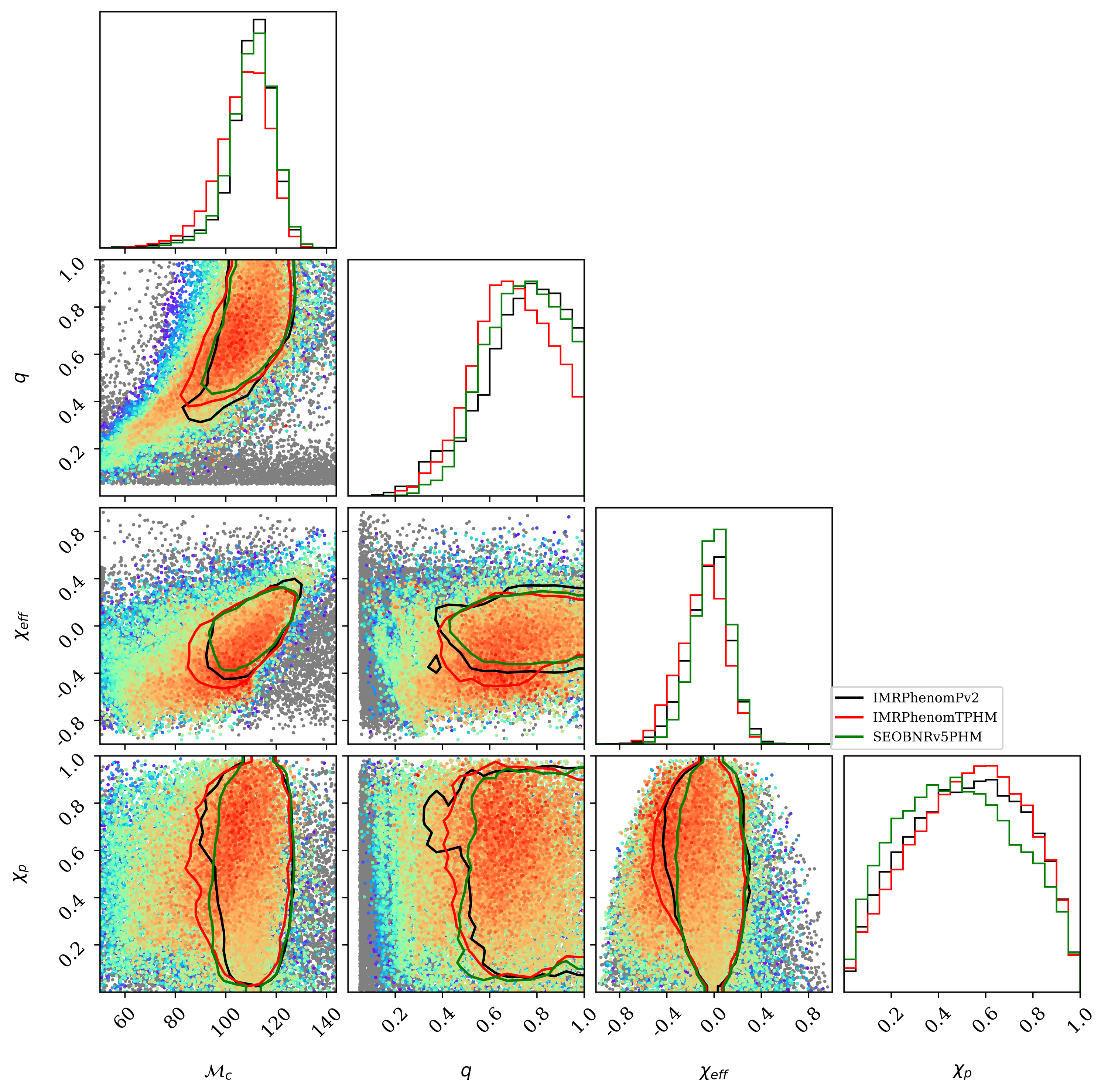}
    \hfill
    \includegraphics[width=0.32\linewidth]{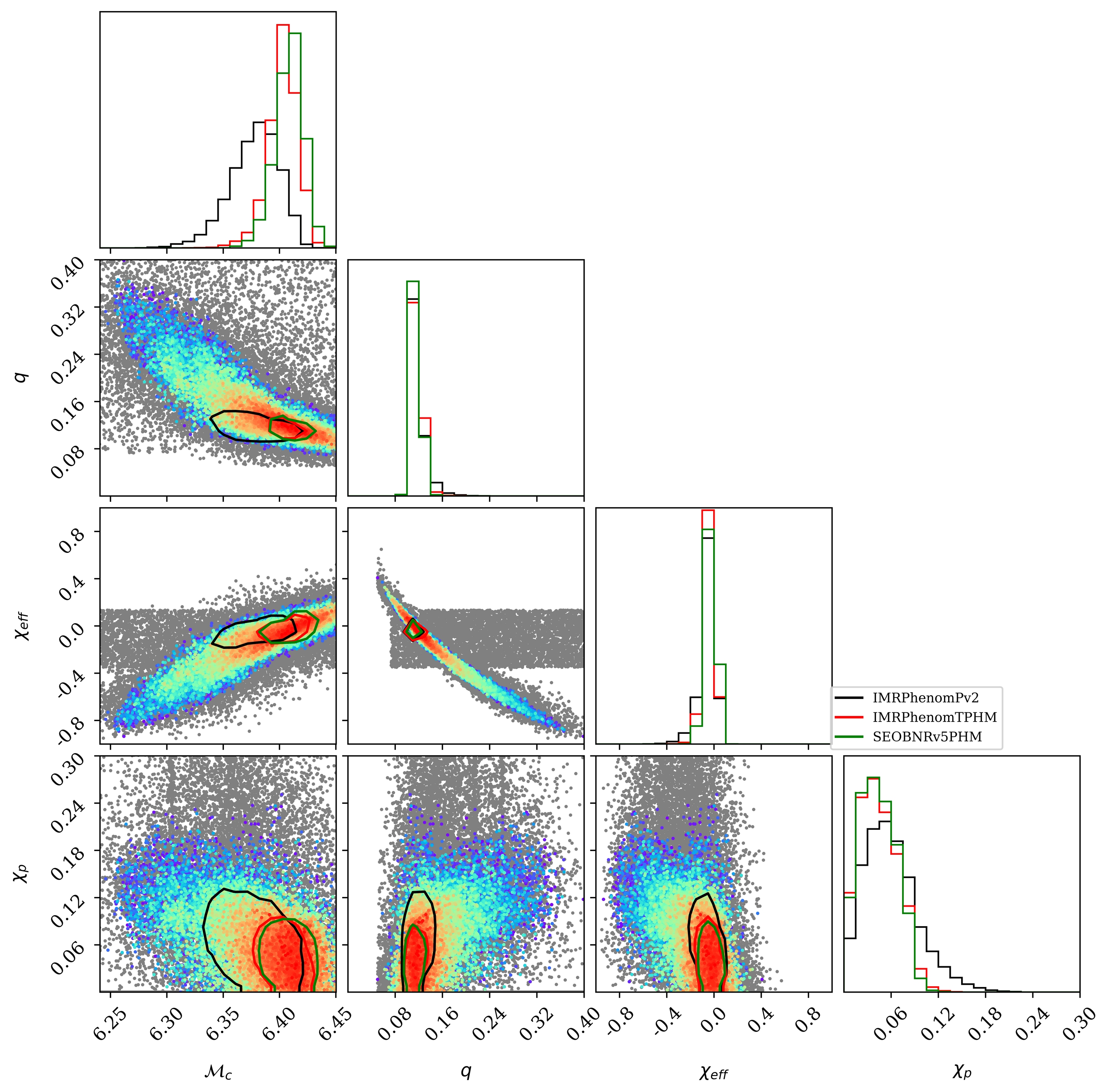}
    \caption[Representative corner plots for three within-work outliers.]{Representative corner plots for three
    within-work outliers. Each panel overlays marginalized posteriors from IMRPhenomPv2, IMRPhenomTPHM, and
    SEOBNRv5PHM, while showing likelihood samples from one state-of-the-art analysis per event: SEOBNRv5PHM for
    GW190412\_053044 (left), IMRPhenomTPHM for GW190521\_030229 (center), and SEOBNRv5PHM for
    GW190814\_211039 (right). The parameters are chirp mass $M_c$, mass ratio $q$, effective spin $\chi_{\rm eff}$,
    and effective in-plane spin $\chi_p$. Contours enclose 90\% credible regions; likelihood samples are colored from
    lower likelihood (gray) to higher likelihood (red). The panels are illustrative diagnostics rather than exhaustive
    model-by-model comparisons, which are available in Ref.~\cite{Manning_2026}.}
    \label{fig:corners_inhouse_representative}
\end{figure*}

These event-level examples revisit comparatively well-understood ground: waveform dependence in GW190412\_053044,
GW190521\_030229, and GW190814\_211039 was already examined in dedicated LIGO--Virgo studies
\cite{LIGO-O3-GW190412,LIGO-O3-GW190521-implications,2020ApJ...896L..44A}. We therefore show only one representative likelihood-sample
layer for each event in Figure \ref{fig:corners_inhouse_representative}, alternating between the two state-of-the-art
waveform families, before turning to the LVK cross-comparisons. These samples provide a compact check that the visible
posterior structure is supported by the explored likelihood surface.

\section{Comparison to and discussion of previous LVK results}
Overall, our reanalyses are broadly congruent with previously published results, with a few notable and interpretable
exceptions. In this section, we systematically document their similarity using JS divergences, identify and discuss
exceptional outliers, and describe a configuration error in several published LVK analyses performed with SEOBNRv4PHM.

These comparisons necessarily include code and settings differences in addition to waveform differences. In particular,
we use a more recent version of RIFT with improved sampling settings, independently generated PSD estimates, and
different sampling rates. Small residual differences from the published posteriors are therefore expected and should not
automatically be attributed to waveform systematics.

The public SEOBNRv4PHM analyses considered here were performed with an earlier version of RIFT, except for the GWTC-2
results for GW190412\_053044, GW190425\_081805, and GW190814\_211039, which used Parallel Bilby in their respective discovery analyses
\cite{LIGO-O3-O3a-catalog,LIGO-O3-GW190412,2020ApJ...892L...3A,2020ApJ...896L..44A}.

\subsection{Similarity of LVK results to reanalyses: JS divergences}

\begin{figure*}
    \centering
    \begin{subfigure}[b]{0.49\textwidth}
        \centering
        \includegraphics[width=\linewidth]{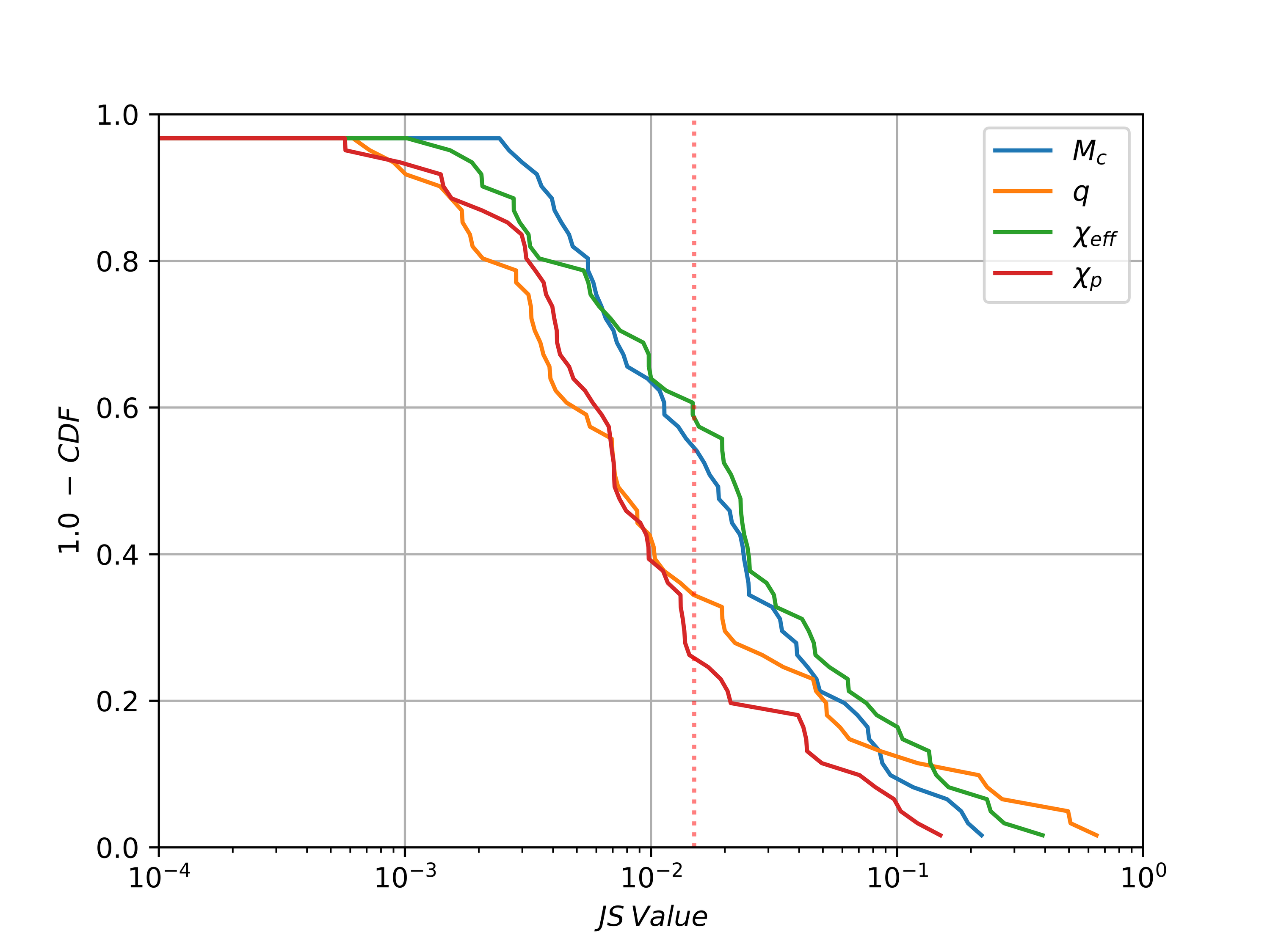}
        \caption{SEOBNRv4PHM vs SEOBNRv5PHM}
        \label{fig:CDFvV}
    \end{subfigure}
    \hfill
    \begin{subfigure}[b]{0.49\textwidth}
        \centering
        \includegraphics[width=\linewidth]{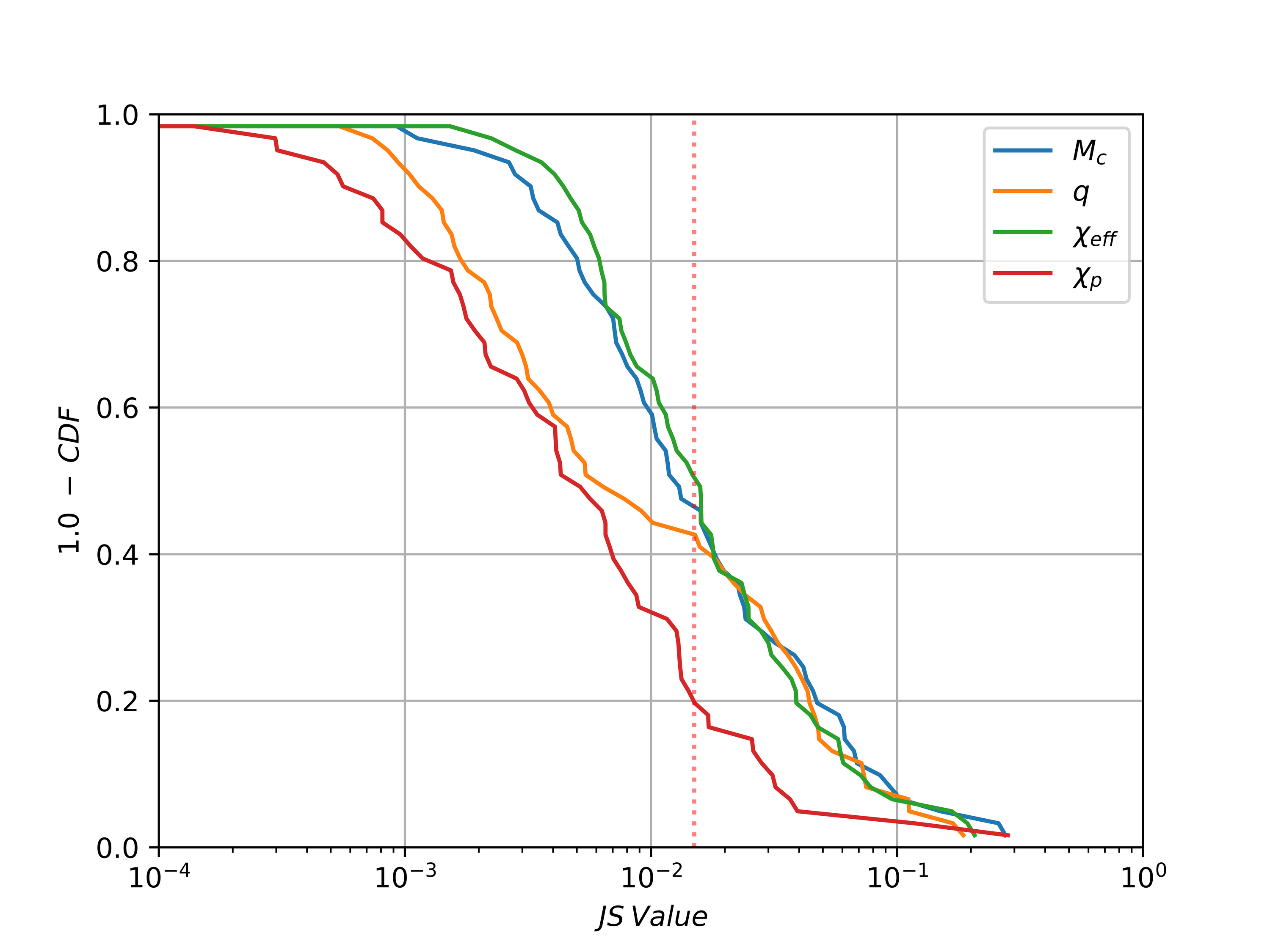}
        \caption{IMRPhenomXPHM vs IMRPhenomTPHM}
        \label{fig:CDFXT}
    \end{subfigure}
    \caption{Inverse CDFs of the Jensen-Shannon Divergence between the most recent GWOSC public analysis and our state-of-the-art model counterpart whose posteriors were generated in this work, for the source frame chirp mass $M_{c}$, mass ratio $q$, effective inspiral spin parameter $\chi_{eff}$, and spin precession parameter $\chi_{p}$ over all events analyzed in this work. A significant fraction of events show noteworthy disagreement $>10^{-2}$ on one or more one-dimensional marginal distributions.}
    \label{fig:GWOSC_CDFs}
\end{figure*}

Figures \ref{fig:CDFXT} and \ref{fig:CDFvV} compare our IMRPhenomTPHM and SEOBNRv5PHM posteriors with the public
IMRPhenomXPHM and SEOBNRv4PHM results, respectively. The contemporary models are broadly congruent with the published
results, but a few events have substantial differences. These outliers span a wide range of masses and signal-to-noise
ratios: they include modest- and low-mass events such as GW191109\_010717 and GW191219\_163120, as well as both loud signals such as
GW200129\_065458 and lower-SNR signals such as GW200208\_130117. Model and analysis systematics therefore occur across the
gravitational-wave census and are not simply predicted by signal strength or binary parameters, consistent with previous
work \cite{gwastro-mergers-TousifGWTC3,gwastro-RIFT-systematics-AnjaliAasim-2020}. For GW200129\_065458 in particular, targeted
LVK-member studies reached different assessments of the evidence for precession after accounting for waveform and
glitch-mitigation uncertainties \cite{2022Natur.610..652H,2022PhRvD.106j4017P}.

About 50\% of the events have at least one parameter
above ${\rm JSD}=0.015$ in both same-family comparisons, corresponding to roughly 30 events. This fraction should not be
read as a pure waveform-error rate because the public and present analyses do not share identical code, PSDs, or
settings. The largest differences, however, are visibly structured and motivate the detailed checks below.

\subsection{Extreme outliers versus LVK results for SEOBNRv4PHM}
\begin{figure*}[h]

  \includegraphics[width=\linewidth]{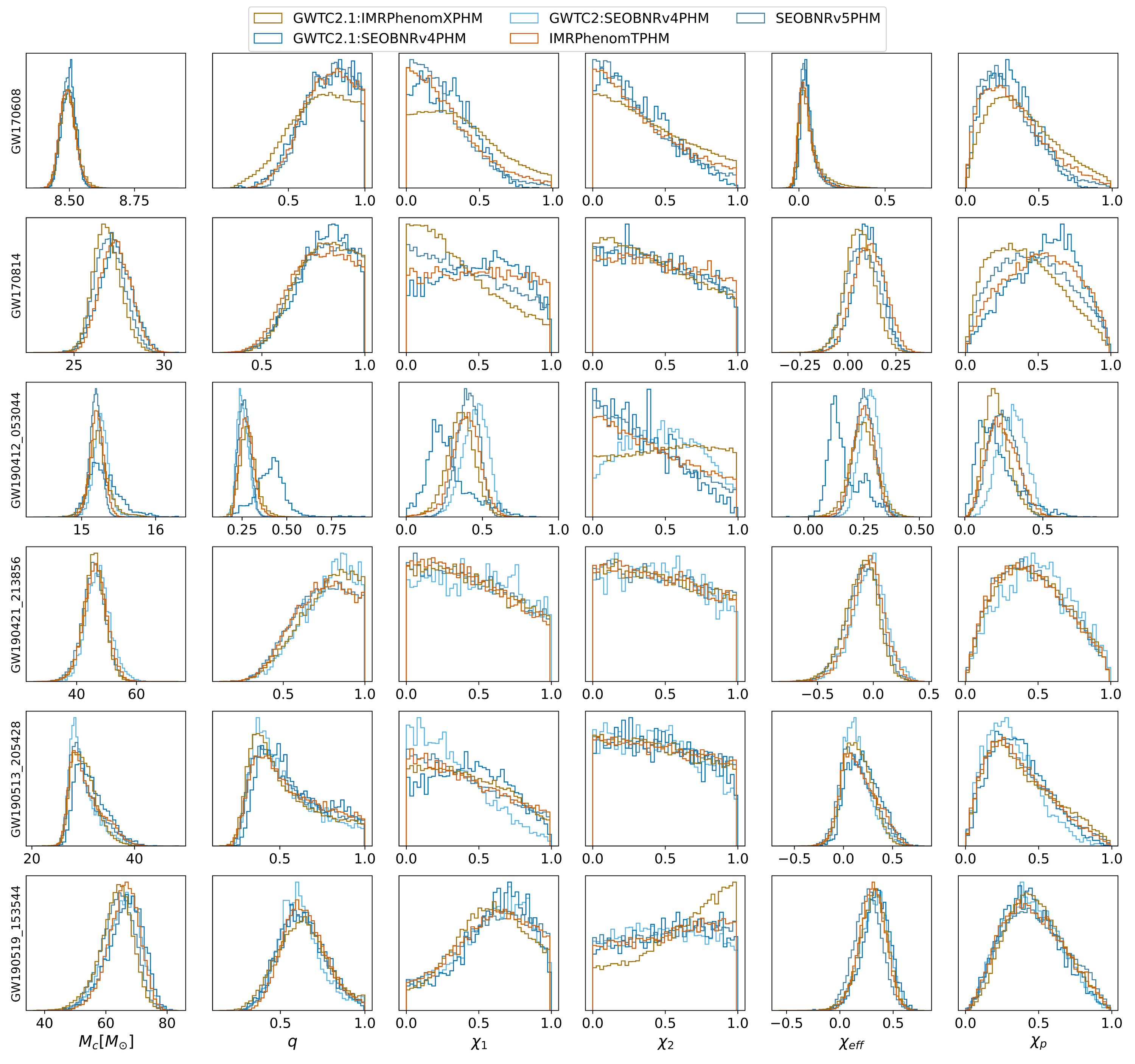}

\caption[Overlapping one-dimensional posterior plots over specific intrinsic parameters to showcase events with the most significant differences.]{ \label{fig:posteriors_gwosc} Posteriors for the chirp mass $M_{c}$, mass ratio $q$, spin magnitudes $\chi_{1}$ and $\chi_{2}$, effective inspiral spin parameter $\chi_{eff}$, and spin precession parameter $\chi_{p}$ for selected events that have the most significant differences between results obtained using SEOBNRv4PHM (light blue), SEOBNRv5PHM (dark blue), IMRPhenomXPHM (light orange), and IMRPhenomTPHM (dark orange).}
\end{figure*}

Figure \ref{fig:posteriors_gwosc} shows selected events with ${\rm JSD}\geq0.015$ between our state-of-the-art analyses
and their previous-generation counterparts. GW190412\_053044 again illustrates that newer models can differ both from one
another and from older models, reinforcing the need for higher-order modes to extract the information in the signal
\cite{LIGO-O3-GW190412}.

For GW190930\_133541, only the newer analyses provide strong support for an unequal-mass system. For GW200224\_222234, only the newer
models recover bimodal structure in the chirp-mass and mass-ratio posteriors. These examples combine waveform evolution
with improvements in parameter-space coverage and convergence, and therefore require direct inspection rather than an
interpretation based on JS divergence alone.

Figure \ref{fig:corners_lvk_crosscomparison} shows the intrinsic-parameter corner plots for GW190930\_133541 and GW200224\_222234
side-by-side. The individual likelihood samples help diagnose whether abrupt posterior features reflect the likelihood
surface or incomplete exploration of parameter space.

\addtocounter{figure}{-1}

\begin{figure*}[h]
\includegraphics[width=\linewidth]{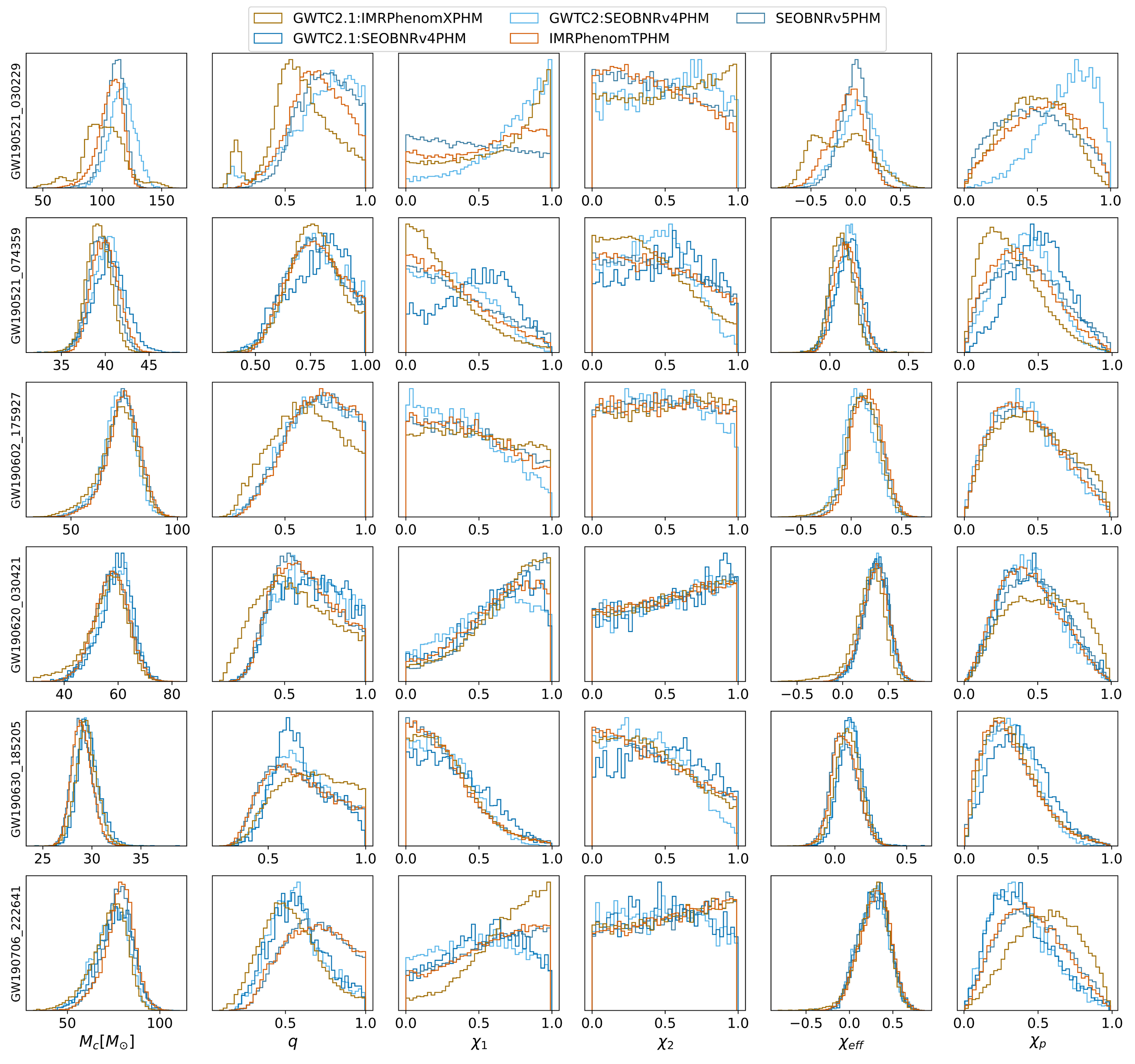}
\caption[]{ \label{fig:posteriors_gwosc_2} (cont.) Posteriors for the chirp mass $M_{c}$, mass ratio $q$, spin magnitudes $\chi_{1}$ and $\chi_{2}$, effective inspiral spin parameter $\chi_{eff}$, and spin precession parameter $\chi_{p}$ for selected events that have the most significant differences between results obtained using SEOBNRv4PHM (light blue), SEOBNRv5PHM (dark blue), IMRPhenomXPHM (light orange), and IMRPhenomTPHM (dark orange).}
\end{figure*}
\addtocounter{figure}{-1}
\begin{figure*}[h]

  \includegraphics[width=\linewidth]{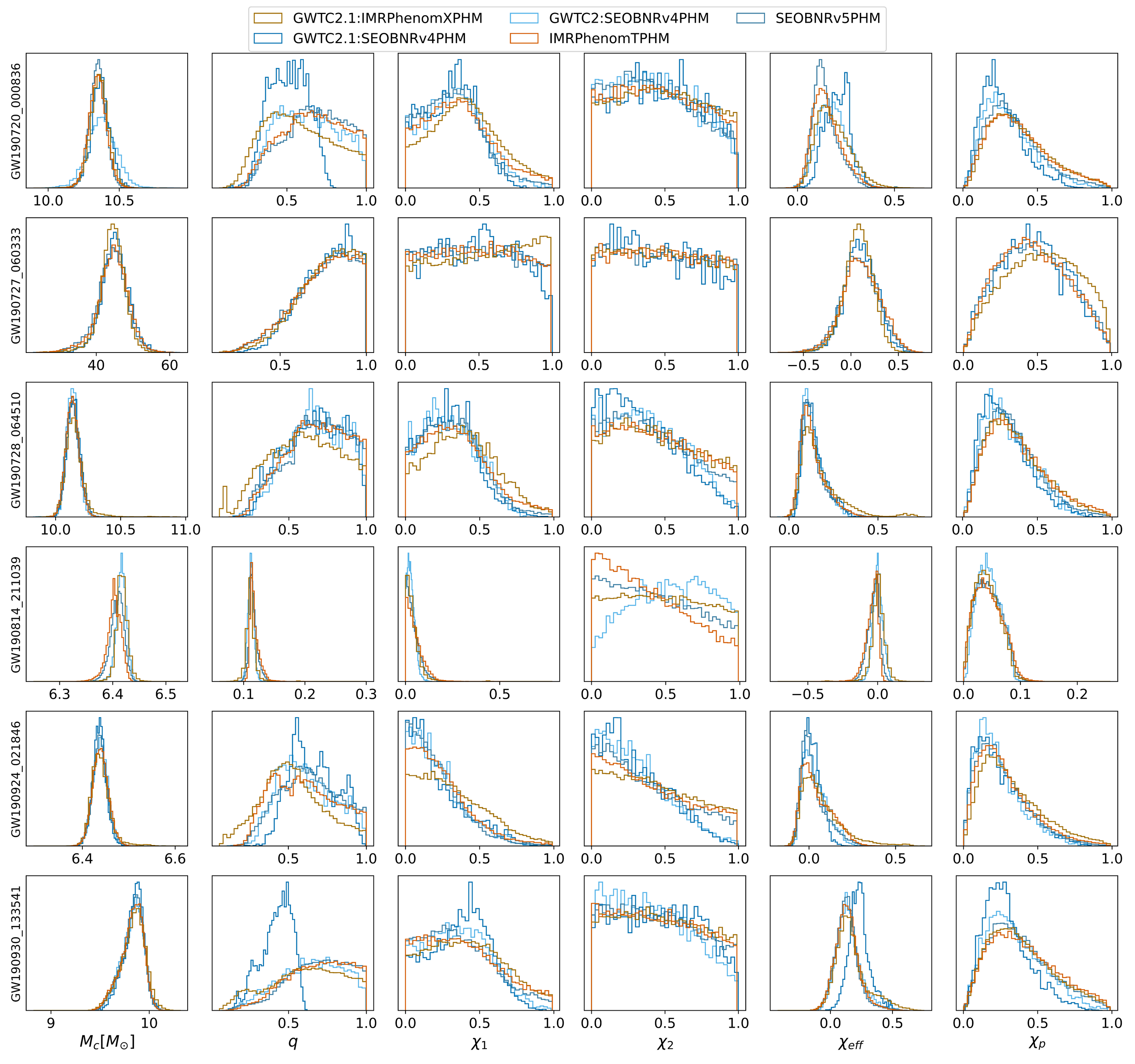}

\caption[]{ \label{fig:posteriors_part_3_gw} (cont.) Posteriors for the chirp mass $M_{c}$, mass ratio $q$, spin magnitudes $\chi_{1}$ and $\chi_{2}$, effective inspiral spin parameter $\chi_{eff}$, and spin precession parameter $\chi_{p}$ for selected events that have the most significant differences between results obtained using SEOBNRv4PHM (light blue), SEOBNRv5PHM (dark blue), IMRPhenomXPHM (light orange), and IMRPhenomTPHM (dark orange).}

\end{figure*}
\addtocounter{figure}{-1}
\begin{figure*}[h]

  \includegraphics[width=\linewidth]{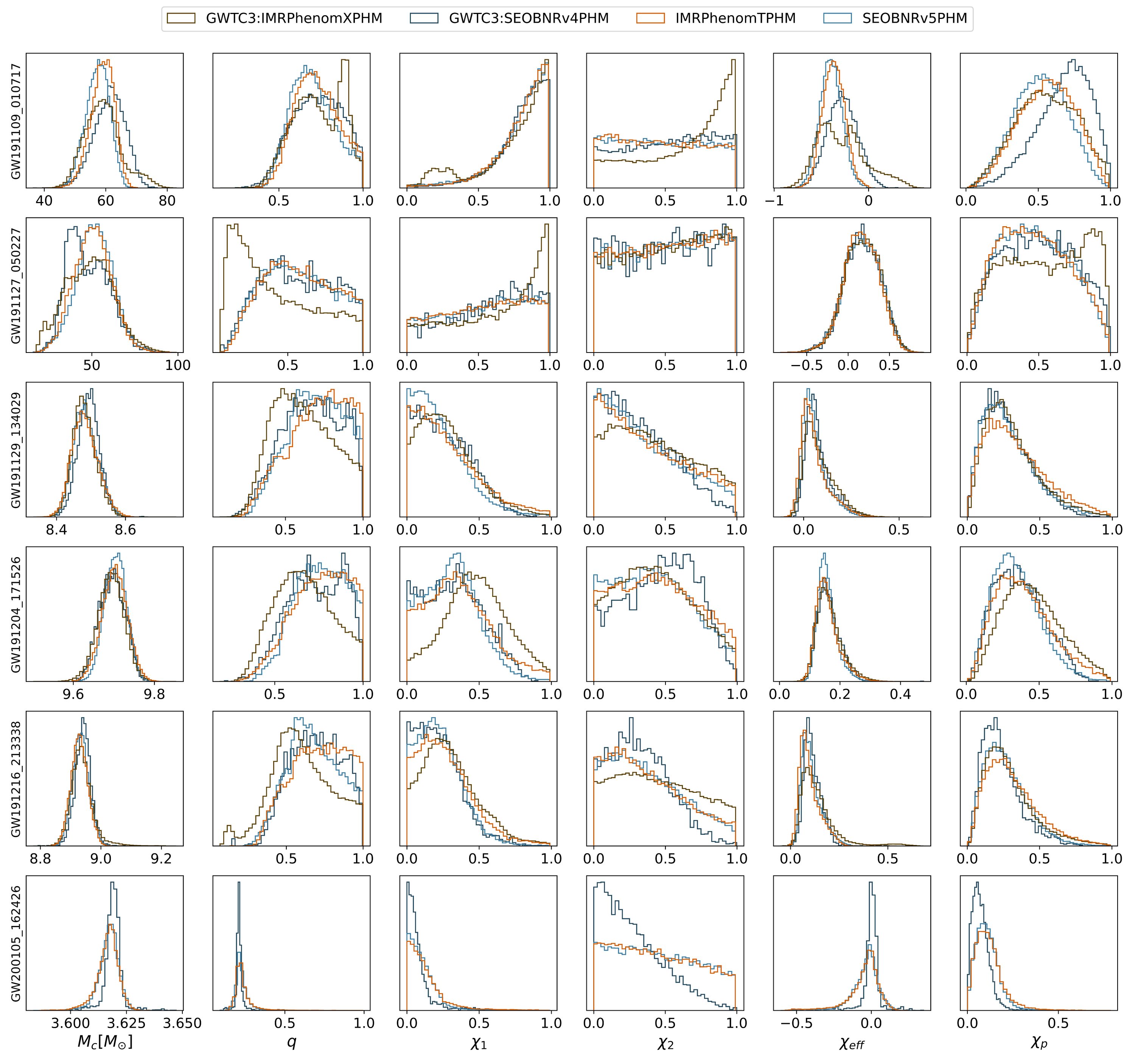}

\caption[]{ \label{fig:posteriors_part_4_gw} (cont.) Posteriors for the chirp mass $M_{c}$, mass ratio $q$, spin magnitudes $\chi_{1}$ and $\chi_{2}$, effective inspiral spin parameter $\chi_{eff}$, and spin precession parameter $\chi_{p}$ for selected events that have the most significant differences between results obtained using SEOBNRv4PHM (light blue), SEOBNRv5PHM (dark blue), IMRPhenomXPHM (light orange), and IMRPhenomTPHM (dark orange).}

\end{figure*}
\addtocounter{figure}{-1}
\begin{figure*}[h]

  \includegraphics[width=\linewidth]{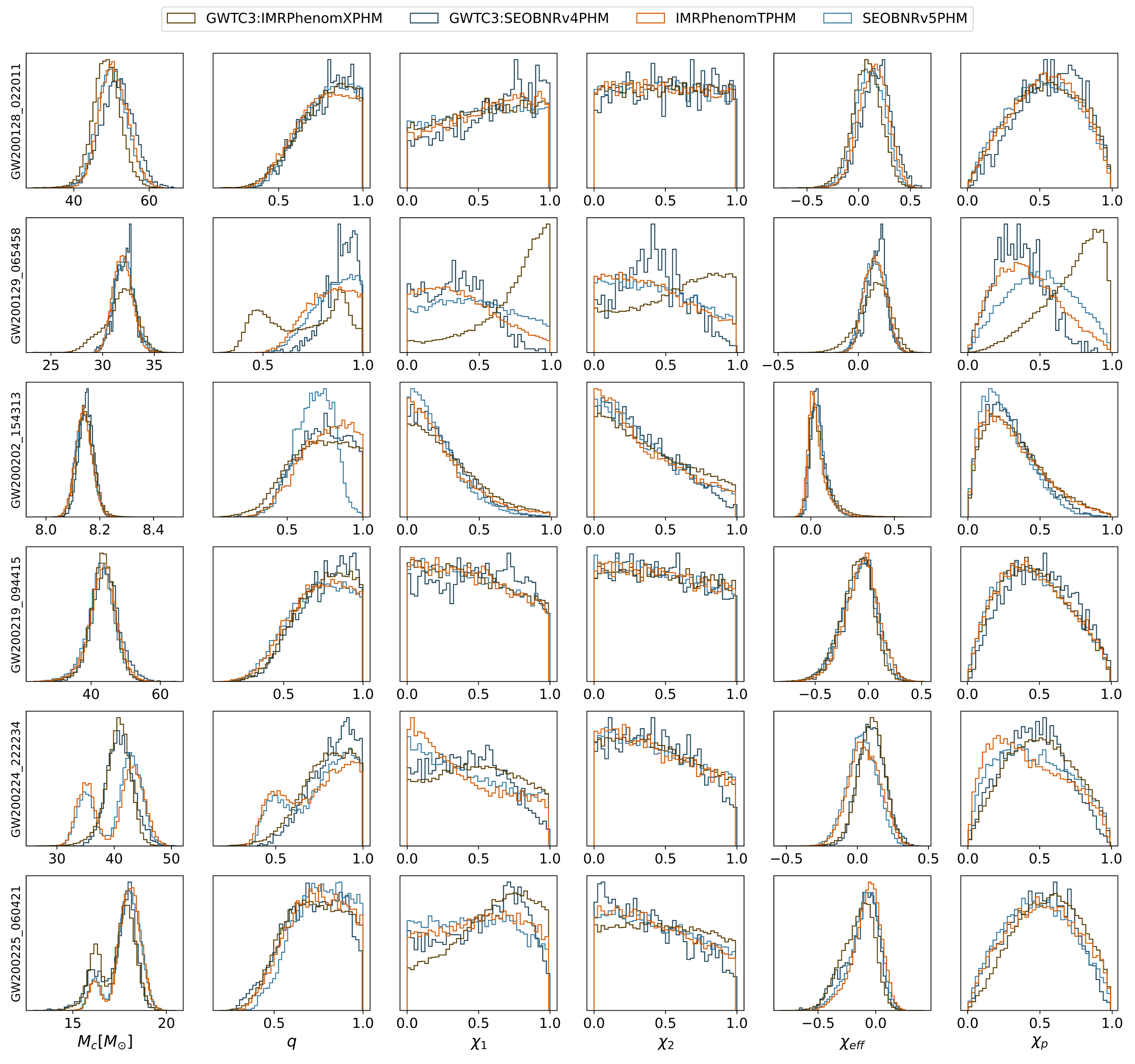}

\caption[]{ \label{fig:posteriors_part_5_gw} (cont.) Posteriors for the chirp mass $M_{c}$, mass ratio $q$, spin magnitudes $\chi_{1}$ and $\chi_{2}$, effective inspiral spin parameter $\chi_{eff}$, and spin precession parameter $\chi_{p}$ for selected events that have the most significant differences between results obtained using SEOBNRv4PHM (light blue), SEOBNRv5PHM (dark blue), IMRPhenomXPHM (light orange), and IMRPhenomTPHM (dark orange).}

\end{figure*}

\begin{figure*}[htb]
    \centering
    \includegraphics[width=0.49\linewidth]{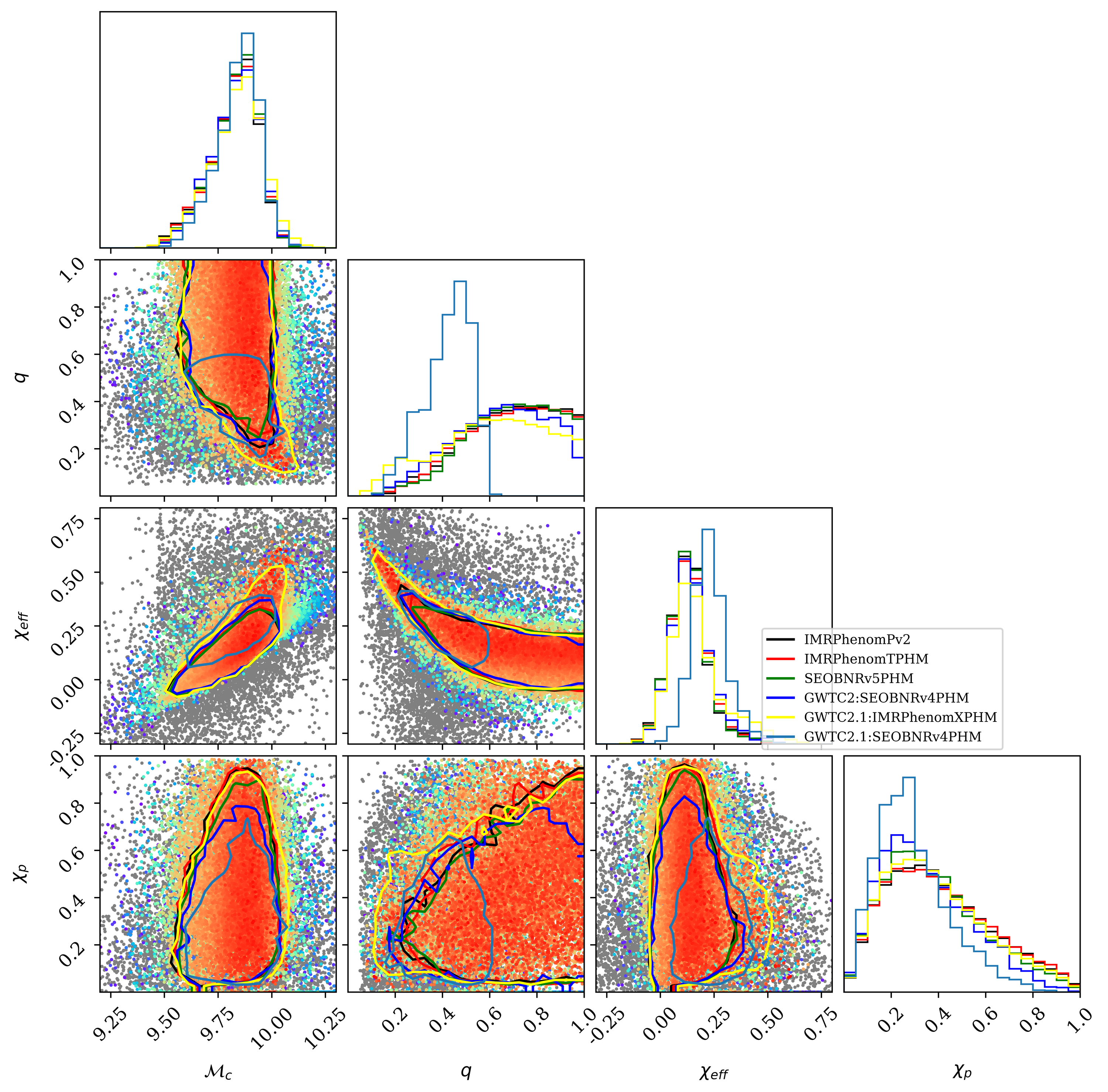}
    \hfill
    \includegraphics[width=0.49\linewidth]{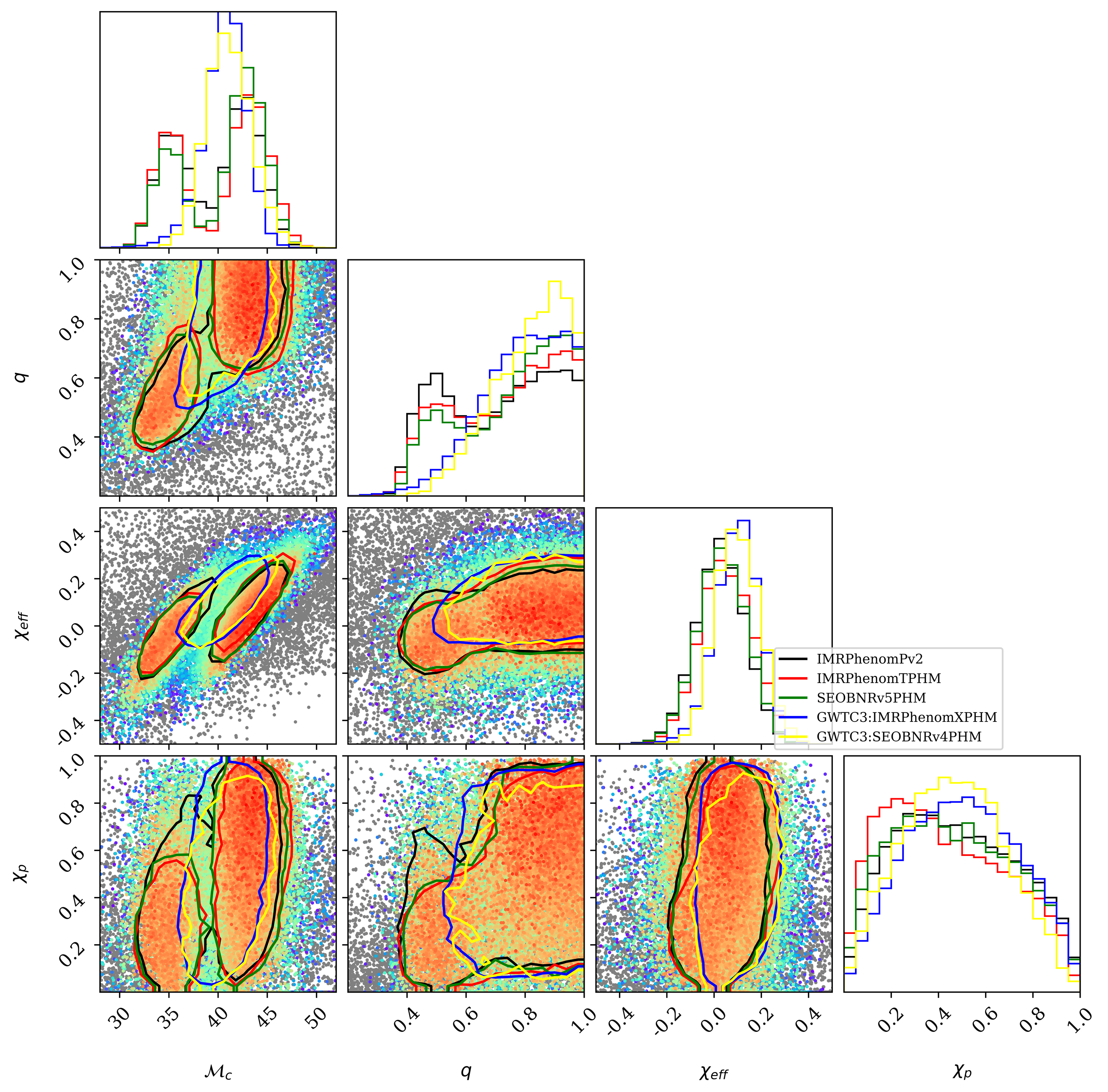}
    \caption[Intrinsic-parameter cross-comparisons for GW190930 and GW200224.]{Intrinsic-parameter corner plots for
    GW190930\_133541 (left) and GW200224\_222234 (right), comparing the published LVK posteriors with the corresponding
    analyses performed in this work. Each panel shows marginalized one-dimensional posteriors and two-dimensional joint
    distributions for the chirp mass $M_c$, mass ratio $q$, effective spin $\chi_{\rm eff}$, and effective in-plane spin
    $\chi_p$. Contours enclose 90\% credible regions; individual likelihood samples are colored from lower likelihood
    (gray) to higher likelihood (red).}
    \label{fig:corners_lvk_crosscomparison}
\end{figure*}

\subsection{Configuration error in published GWTC-2.1 analyses and a later reprocessing}

\begin{figure*}[htb]
    \centering
    \begin{subfigure}[b]{0.49\textwidth}
        \centering
        \includegraphics[width=\linewidth]{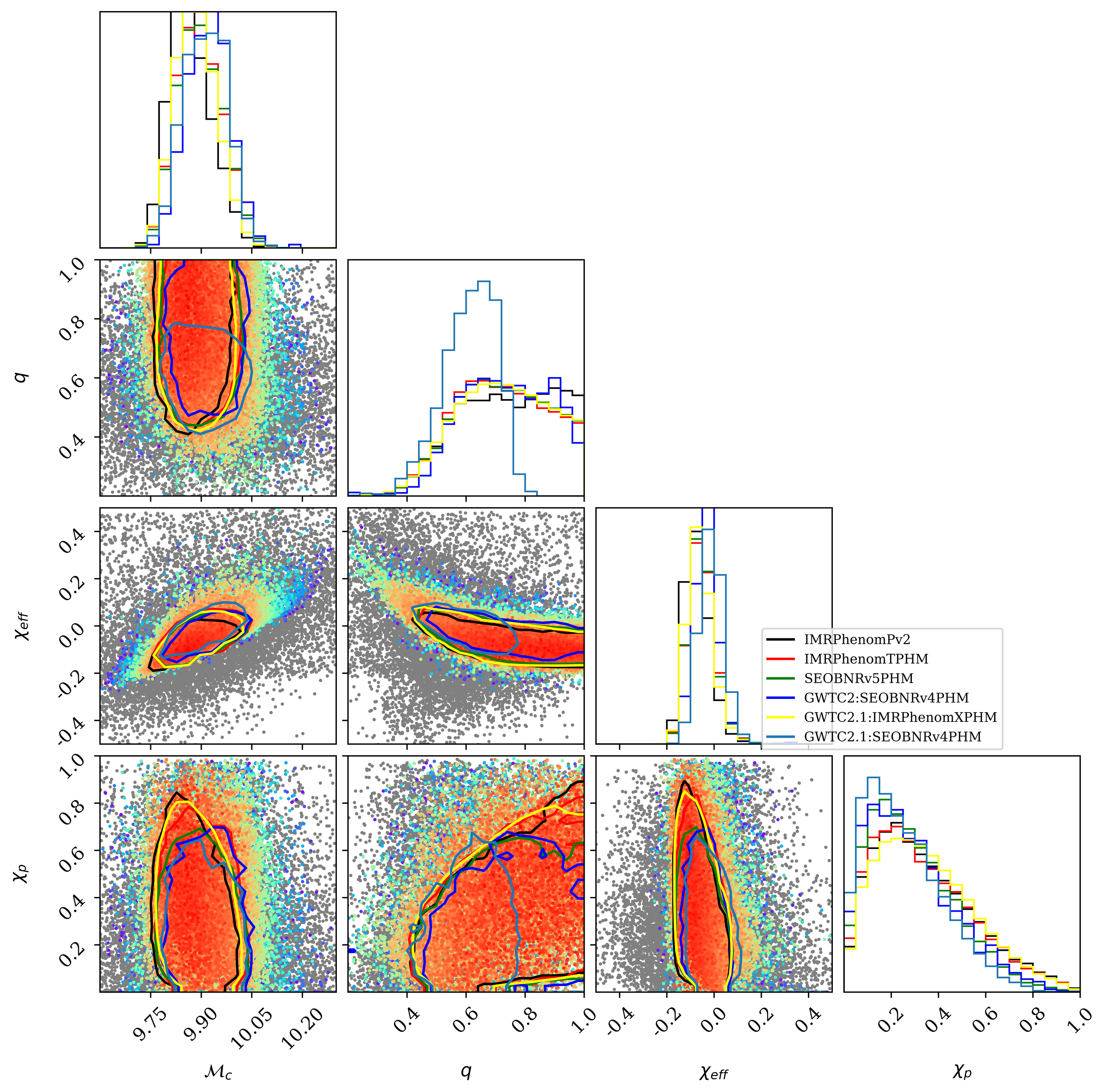}
        \caption{GW190707\_093326}
        \label{fig:190707_gwosc}
    \end{subfigure}
    \hfill
    \begin{subfigure}[b]{0.49\textwidth}
        \centering
        \includegraphics[width=\linewidth]{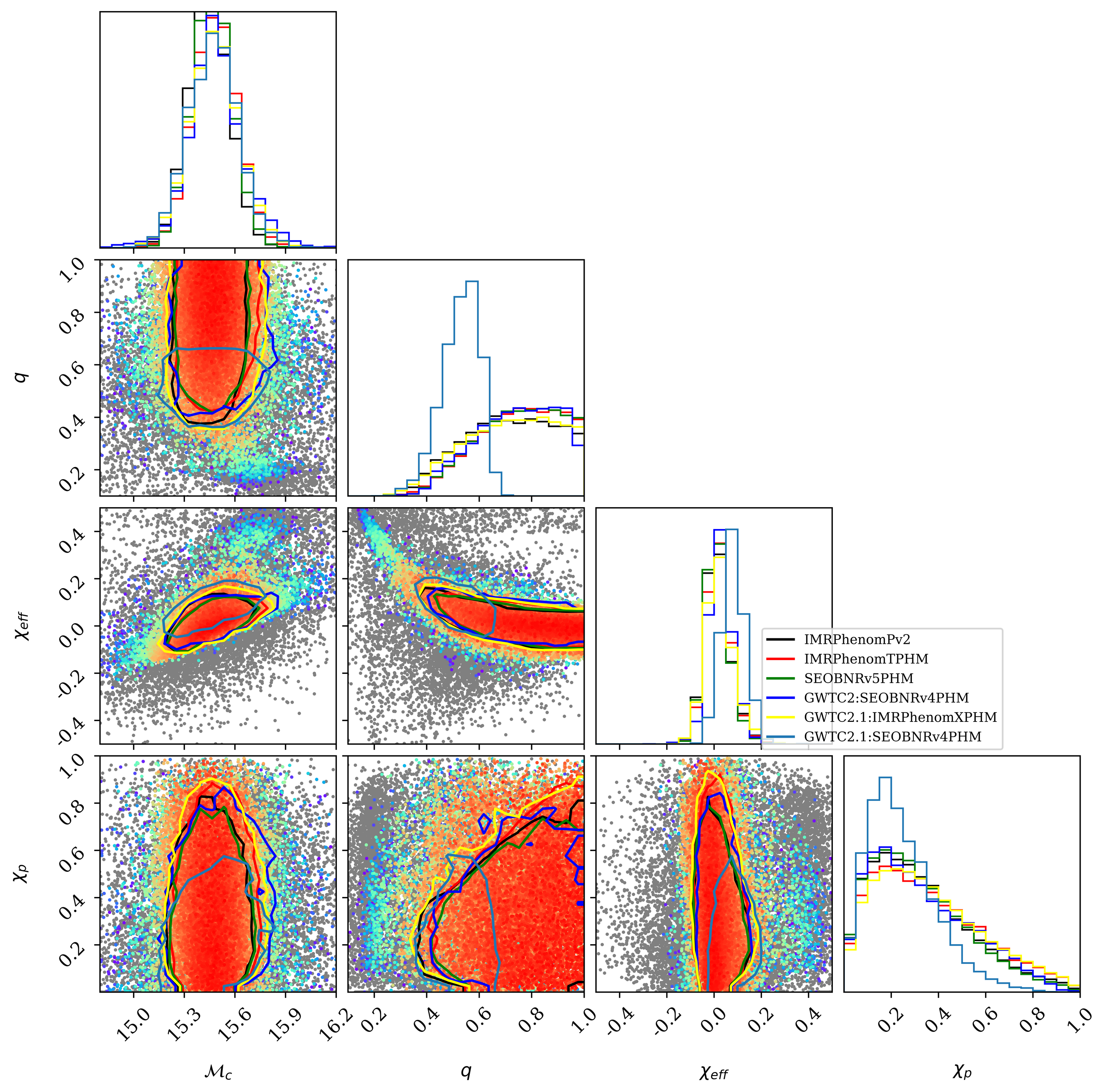}
        \caption{GW190708\_232457}
        \label{fig:190708_gwosc}
    \end{subfigure}

    \vspace{1em}

    \begin{subfigure}[b]{0.49\textwidth}
        \centering
        \includegraphics[width=\linewidth]{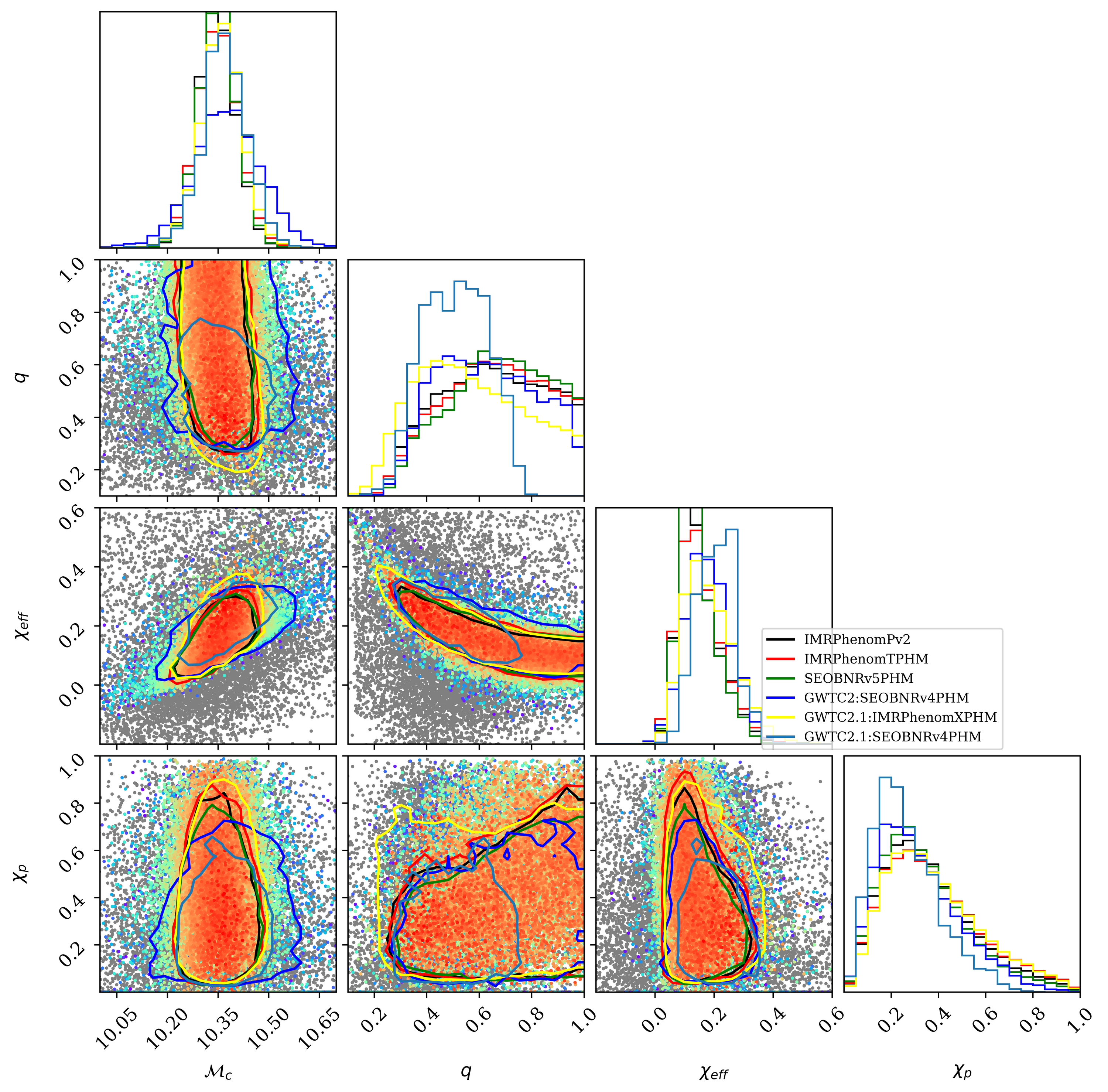}
        \caption{GW190720\_000836}
        \label{fig:190720_gwosc}
    \end{subfigure}
    \hfill
    \begin{subfigure}[b]{0.49\textwidth}
        \centering
        \includegraphics[width=\linewidth]{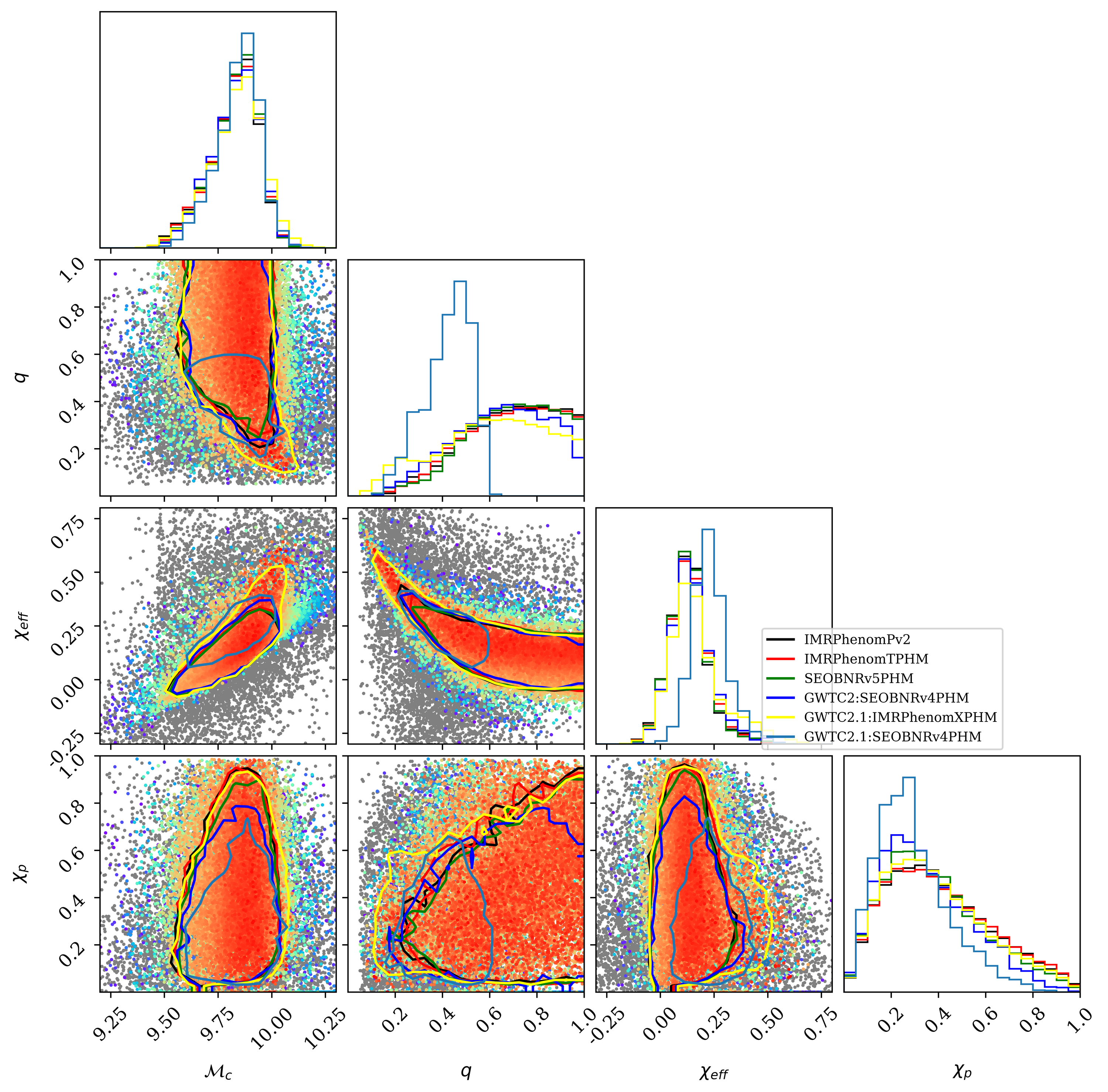}
        \caption{GW190930\_133541}
        \label{fig:190930_gwosc}
    \end{subfigure}
    \caption[Anecdotal corner plots for events whose published SEOBNRv4PHM analyses show railing in mass ratio.]{Corner plots showing the posterior distribution for events whose published SEOBNRv4PHM analyses show railing in mass ratio, including marginalized one-dimensional posteriors and two-dimensional joint distributions for the chirp mass $M_{c}$, mass ratio $q$, effective spin $\chi_{eff}$, and effective in-plane spin $\chi_{p}$. The lined contours correspond to the 90\% credible intervals, while the colored points represent individual likelihood samples.}
    \label{fig:corners_gwosc}
\end{figure*}

\noindent \emph{Inadvertently over-consistent settings}:
SEOBNRv4PHM does not allow the end user to generate a waveform in regions where any mode of the waveform would have
frequency content above the Nyquist frequency of the $(\ell_{\rm max},\ell_{\rm max})$ mode for $\ell_{\rm max}=4$.  As a result, at a fixed sampling rate, for sufficiently low mass binaries, some parts of the
parameter space cannot be explored at all, producing artificial features in the posterior.

By contrast, the newer SEOBNRv5PHM model allows end users to mitigate this limitation by selecting the $\ell_{\rm max}$
which causes waveform generation to fail. In our newer analyses, we adopt $\ell_{\rm max}=1$ where needed, bypassing
this check and allowing SEOBNRv5PHM to cover a larger fraction of the parameter space at the same sampling rate as the
other waveforms.

Published LVK GWTC-2.1 analyses adopted consistent PSD and sampling-rate settings across these events
\cite{LIGO-O3-O3a_final-catalog,GWTC21-PEDataRelease}. Unfortunately, this configuration produces adverse results for
six low-mass GWTC-2.1 event reanalyses---GW190707\_093326, GW190708\_232457, GW190720\_000836,
GW190728\_064510, GW190924\_021846, and GW190930\_133541---for which the SEOBNRv4PHM Nyquist-frequency
limitation artificially truncates the posterior away from $q=1$. The same problem appears in the later public
reprocessing of GW200105\_162426. Specifically, the affected posterior is the one distributed in version 2 of the
public parameter-estimation data release accompanying the later catalog paper, published on 23 October 2023
\cite{LIGO-O3-O3b-catalog,GWTC3-PEDataRelease}. By contrast, the original dedicated GW200105 analysis and its
corresponding posterior-sample release, LIGO-P2100143-v2, were correct and did not exhibit this problem
\cite{LIGO-O3-NSBH,GW200105-Original-PEDataRelease}.

Figure \ref{fig:corners_gwosc} shows the posteriors for four of these events.

Garr\'on and Keitel previously identified several apparent waveform discrepancies in published RIFT results as
sampling or convergence problems \cite{GarronKeitel2024}. Their SEOBNRv4PHM discrepancy for the low-mass pair
GW190707\_093326--GW190930\_133541 uses two of the affected GWTC-2.1 results above; its mass-posterior symptom is
attributable to the same Nyquist-frequency configuration mechanism identified here. Their other examples, including a
missing NRSur7dq4 sky mode for GW190719\_215514 and higher-mass RIFT outliers, are distinct sampling/convergence failures
rather than consequences of this SEOBNRv4PHM waveform-domain restriction; Wagner et al. discuss these convergence
failures in detail and show how they can be resolved with improved RIFT settings \cite{gwastro-RIFT_FinerNet}.

\subsection{Interpretation and limitations}

Our comparisons involve five waveform models, but only IMRPhenomPv2, IMRPhenomTPHM, and SEOBNRv5PHM were analyzed
within the same framework. The public IMRPhenomXPHM and SEOBNRv4PHM results were produced with different code versions,
PSDs, sampling rates, and other event-specific settings. Minor differences between our results and the public posteriors
are therefore expected and cannot be assigned uniquely to waveform physics.

This limitation does not affect the direct comparison between SEOBNRv5PHM and IMRPhenomTPHM, for which the analysis
framework is controlled and overall agreement is good. Nor does it explain the artificial mass-ratio boundaries in the
six GWTC-2.1 SEOBNRv4PHM results or the later GW200105 reprocessing, whose locations follow directly from the model's
Nyquist-frequency check at the adopted sampling rates. More generally, the outliers identified here show why
posterior-level comparisons and inspection of sampling diagnostics remain necessary even when catalog-level agreement
is strong.

\section{Conclusions}
\label{sec:conclude}

In this paper, we use multiple waveform models to reanalyze 60 GWTC-3 events with consistent analysis settings.
Combining the asimov production-quality inference automation with the RIFT distributed parameter-inference code, we
perform analyses at scale without requiring access to special-purpose data resources. Our controlled set compares two
time-domain models with higher-order modes against one frequency-domain model without higher-order modes.

We find that the two state-of-the-art models, SEOBNRv5PHM and IMRPhenomTPHM, largely agree: about 20\% of events have a
visible difference in at least one one-dimensional marginal posterior. These outliers demonstrate the continuing value
of analyses with multiple waveform families under consistent settings and within a common code base.

Our comparison with the older IMRPhenomPv2 model exemplifies the improvements made in gravitational-wave models over
the past decade. Although occasionally adequate, IMRPhenomPv2 often infers qualitatively different binary parameters
from both state-of-the-art models across the gravitational-wave census.

Finally, comparison with published LVK results reveals a bug in six low-mass GWTC-2.1 SEOBNRv4PHM results and in a
later reprocessing of GW200105. At the adopted sampling rates, the model's Nyquist-frequency restriction prevented
exploration of part of the allowed parameter space, artificially truncating the mass-ratio posterior away from $q=1$
and producing misleading results.

By analyzing 60 publicly available events released across GWTC-1, GWTC-2, GWTC-2.1, and GWTC-3 with
state-of-the-art models in a consistent framework, we
establish a benchmark for parameter estimation with the RIFT pipeline.

\clearpage
\begin{acknowledgments}
  ROS acknowledges support from NSF PHY 2012057, NSF PHY 2309172 and 2206321. The authors are grateful for computational resources provided by the LIGO Laboratory and supported by National Science Foundation Grants PHY-0757058 and PHY-0823459; no other institutional computing resources were used. This research has made use of data or software obtained from the Gravitational Wave Open Science Center (gwosc.org), a service of the LIGO Scientific Collaboration, the Virgo Collaboration, and KAGRA. This material is based upon work supported by NSF's LIGO Laboratory which is a major facility fully funded by the National Science Foundation, as well as the Science and Technology Facilities Council (STFC) of the United Kingdom, the Max-Planck-Society (MPS), and the State of Niedersachsen/Germany for support of the construction of Advanced LIGO and construction and operation of the GEO600 detector. Additional support for Advanced LIGO was provided by the Australian Research Council. Virgo is funded, through the European Gravitational Observatory (EGO), by the French Centre National de Recherche Scientifique (CNRS), the Italian Istituto Nazionale di Fisica Nucleare (INFN) and the Dutch Nikhef, with contributions by institutions from Belgium, Germany, Greece, Hungary, Ireland, Japan, Monaco, Poland, Portugal, Spain. KAGRA is supported by Ministry of Education, Culture, Sports, Science and Technology (MEXT), Japan Society for the Promotion of Science (JSPS) in Japan; National Research Foundation (NRF) and Ministry of Science and ICT (MSIT) in Korea; Academia Sinica (AS) and National Science and Technology Council (NSTC) in Taiwan.
\end{acknowledgments}

\input{paper.bbl}

\end{document}